\documentclass[hidelinks]{article}

\usepackage{arxiv}
\usepackage{subcaption} 
\usepackage[numbers]{natbib}
\usepackage[utf8]{inputenc} 
\usepackage[T1]{fontenc}    
\usepackage{hyperref}       
\usepackage{url}            
\usepackage{booktabs}       
\usepackage{amsfonts}       
\usepackage{nicefrac}       
\usepackage{microtype}      
\usepackage{lipsum}		
\usepackage{graphicx}
\usepackage{natbib}
\usepackage{doi}
\usepackage{amsmath}
\usepackage[table]{xcolor}  
\usepackage{booktabs}

\title{Storm Surge in Color: RGB-Encoded Physics-Aware Deep Learning for Storm Surge Forecasting}

\author{
\textbf{
Jinpai Zhao\textsuperscript{1}\thanks{Corresponding author: max.zhao@utexas.edu}
, 
Albert Cerrone\textsuperscript{2}, 
Eirik Valseth\textsuperscript{3,4}, 
Leendert Westerink\textsuperscript{2}, 
Clint Dawson\textsuperscript{1}
} \\
\\
\textsuperscript{1}Oden Institute for Computational Engineering \& Sciences \\ 
University of Texas at Austin, Austin, TX, USA \\
\textsuperscript{2}Department of Civil \& Environmental Engineering \& Earth Sciences \\ 
University of Notre Dame, South Bend, IN, USA \\
\textsuperscript{3}Department of Data Science \\ 
Norwegian University of Life Sciences, Ås, Norway \\
\textsuperscript{4}Department of Numerical Analysis and Scientific Computing  \\ 
Simula Research Laboratory, Oslo, Norway \\
}

\date{}

\renewcommand{\shorttitle}{}

\hypersetup{
pdftitle={Storm surge in color},
pdfsubject={NA,CS},
pdfauthor={Jinpai Zhao, Albert Cerrone, Eirik Valseth, Leendert Westerink, Clint Dawson},
pdfkeywords={RGB, ADCRIC, Surrogate Model, LSTM},
}

\begin{document}
\maketitle

\begin{abstract}
	Storm surge forecasting plays a crucial role in coastal disaster preparedness, yet existing machine learning approaches often suffer from limited spatial resolution, reliance on coastal station data, and poor generalization. Moreover, many prior models operate directly on unstructured spatial data, making them incompatible with modern deep learning architectures. In this work, we introduce a novel approach that projects unstructured water elevation fields onto structured Red Green Blue (RGB)-encoded image representations, enabling the application of Convolutional Long Short Term Memory (ConvLSTM) networks for end-to-end spatiotemporal surge forecasting. Our model further integrates ground-truth wind fields as dynamic conditioning signals and topo-bathymetry as a static input, capturing physically meaningful drivers of surge evolution. Evaluated on a large-scale dataset of synthetic storms in the Gulf of Mexico, our method demonstrates robust 48-hour forecasting performance across multiple regions along the Texas coast and exhibits strong spatial extensibility to other coastal areas. By combining structured representation, physically grounded forcings, and scalable deep learning, this study advances the frontier of storm surge forecasting in usability, adaptability, and interpretability.
\end{abstract}


\section{Introduction}
Storm surge is one of the most devastating consequences of tropical cyclones, with the potential to cause widespread coastal flooding, infrastructure damage, and significant loss of life. Events such as Hurricane Katrina (2005) and Hurricane Ike (2008) have underscored the critical need for accurate, high-resolution storm surge forecasts to support emergency response and long-term risk mitigation. With rising sea levels and an increase in the frequency and intensity of extreme weather events, improving the reliability and scalability of storm surge prediction models remains an urgent priority.

Traditionally, high-fidelity physics-based numerical models such as the  ADvanced CIRCulation model (ADCIRC) \cite{adcirc,adcircv55} and its relatives~\cite{dawson2011discontinuous,wichitrnithed2024discontinuous} have been used to simulate storm surge with remarkable accuracy. These models typically solve the two-dimensional shallow water equations over high resolution unstructured meshes that can represent complex coastal topography and bathymetry. However, this precision comes at the cost of computational expense. While ADCIRC runs exceptionally fast on current-generation HPC hardware, forecasting a single storm is not instantaneous.  For example, on the Texas Advanced Computing Center (TACC) Frontera cluster, the ADCIRC-driven STOFS-2D-Global can render a 7-day forecast in approximately 20 wall-clock minutes over 2400 cores.  In the absence of such resources or if probabilistic guidance is desired, keeping pace with forecast windows would be challenging.

Data-driven modeling has long been explored as an efficient alternative to high-fidelity physics-based solvers for storm surge prediction. One of the earliest efforts, by Lee \cite{lee2009}, applied a back-propagation neural network to forecast storm surge in Taiwan using meteorological and tidal parameters. Building on this foundation, Kajbaf \emph{et al.} in \cite{kajbaf2020} conducted a comprehensive assessment of several surrogate modeling techniques—including artificial neural networks, support vector regression, and Gaussian process regression—for regional surge estimation, emphasizing trade-offs between accuracy and computational cost. More recent works have introduced novel methods to handle the temporal and spatial complexities of storm surge dynamics. In \cite{pachev2023}, Pachev \emph{et al.} proposed a flexible, point-based surrogate framework that aggregates time and space dimensions into a two-stage prediction pipeline for peak surge, while Lockwood \emph{et al.} \cite{lockwood2022} trained individual feedforward neural networks at each time step to estimate surge evolution based on storm characteristics.  Other studies have leveraged spatially-agnostic approaches wherein time-series-forecast ML architectures are used to predict water levels.  For example, Giaremis \emph{et al.} \cite{giaremis2024} used LSTM networks to model and correct systematic bias in physics-based storm tide predictions by learning from past hurricane observations. Further increasing model complexity, Cerrone \emph{et al.} \cite{cerrone2025} applied a Temporal Fusion Transformer (TFT) to NOAA’s operational water level forecasts, demonstrating improved predictive skill over a seven-day horizon at hundreds of observation stations.

While those methods emphasize temporal treatment or decoupled prediction, other studies aim to model spatiotemporal dynamics more explicitly. In \cite{lee2021}, Lee \emph{et al.} introduced a one-dimensional convolutional neural network (C1PKNet) that maps tropical cyclone track time series to peak surge fields across the Chesapeake Bay region. In a recent study, Naeini \emph{et al.} \cite{Naeini2025Advancing} developed a hybrid approach wherein hierarchical deep neural networks (HDNNs) and convolution autoencoders are coupled to predict hurricane-induced storm surge over a spatiotemporal domain.  This hybrid model was trained to mimic ADCIRC, something the modeling strategy detailed herein was compelled to do. These approaches demonstrate strong performance within specific domains. Models trained on single-location or station-level data are particularly effective for site-specific forecasting, and time-segmented methods offer computational benefits for rapid prediction. However, many of these frameworks are limited in their ability to generalize across space and storm types, especially when applied to regions or trajectories not seen during training. Moreover, most models are not designed to recover the full spatiotemporal surge field, which is critical for informing broad-scale emergency response and infrastructure planning.

Some might question pursuing ML-based spatiotemporal techniques in the hydrodynamic space, especially given relatively expedient and accurate forward models such as SLOSH~\cite{jelesnianski1992}, SFINCS~\cite{leijnse2021}, and SCREAM~\cite{donahue2024}.  First, while these models run faster than traditional finite element based circulation models, their run-times are certainly not instantaneous.  While they could be used in a probabilistic setting, their use in non-invasive uncertainty quantification techniques like Markov chain Monte Carlo is limited.  Additionally, these models are not necessarily readily modifiable to include previously unaccounted for physics.  Most ML frameworks, on the other hand, can be easily retrained to consider new species of exogenous data (e.g. hydrologic discharge, sea surface temperature, etc.) while running practically instantaneously on even modest Graphical Processing Units (GPUs).  An easily extendable ML framework that retains the accuracy of performant forward models in the hydrodynamic space would facilitate producing stochastic guidance.

In this work, we propose a novel deep learning framework that addresses these limitations by projecting unstructured water elevation fields onto structured RGB-encoded grids, enabling full-resolution spatiotemporal forecasting using Convolutional LSTM networks. Our model is conditioned on temporally evolving wind fields and incorporates static bathymetry as an auxiliary input channel, allowing it to capture both dynamic forcing and spatial context. This representation facilitates the use of modern, scalable neural architectures while preserving key physical structures. The RGB encoding naturally smooths the output, aligning well with the inductive bias of convolutional networks and enhancing generalization without sacrificing detail. We evaluate our model on a large dataset of synthetic storms along the Texas coast, demonstrating accurate 48-hour predictions across multiple regions, including out-of-distribution generalization to unseen areas, while achieving orders-of-magnitude speedup over traditional numerical solvers. The remainder of this paper is organized as follows: In Section \ref{sec:data_pipeline}, we describe the dataset and processing pipeline. In Section \ref{sec:method}, model architecture and forecasting strategy are outlined, followed by the configuration settings and evaluation metrics in Section \ref{sec:config}. We present the results from our models and discussions in Section \ref{sec:result}. Finally, in  Section \ref{sec:conclusion} we summarize our findings and some potential directions for future work.

\section{Dataset and Processing Pipeline} \label{sec:data_pipeline}

\subsection{Overview of the ADCIRC model and Texas FEMA dataset}

Storm surge forecasting traditionally relies on two types of data sources: observational records from coastal tide gauges, such as those provided by NOAA’s Tides and Currents platform \cite{noaa}, and simulated outputs from high-fidelity numerical models. While gauge observations are invaluable for validating predictions, they are spatially sparse and generally restricted to specific coastal stations, making them unsuitable for learning high-resolution spatiotemporal patterns.

In contrast, high-fidelity numerical models such as ADCIRC  offer dense spatial coverage at millions of computational mesh nodes and enable the simulation of a wide range of hypothetical and historical storm scenarios. ADCIRC has been rigorously validated against real hurricane events, such as Hurricane Katrina (2005), Hurricane Ike (2008), and Hurricane Harvey (2017), and is widely considered a benchmark for coastal storm surge modeling, see e.g., \cite{hope2013, dietrich2012,goff2019}. Its ability to accurately reproduce storm tide, wave, and forerunner behavior across broad coastal domains makes it particularly well-suited for generating training datasets for data-driven models.

ADCIRC solves the two-dimensional, depth-integrated shallow water equations (SWE) in spherical coordinates. These equations are discretized by applying the Bubnov-Galerkin finite element method with linear polynomial basis functions to the weak form of the SWE in the spatial domain and  finite difference schemes for time integration. The use of unstructured triangular meshes allows ADCIRC to resolve complex coastal geometries, estuaries, and floodplains with locally refined resolution. External forcing, such as wind stress and atmospheric pressure, can be applied from synthetic or observational storm profiles, making it flexible for both retrospective studies and probabilistic hazard assessment. The governing equations, in primitive and non-conservative form in spherical coordinates, are as follows \cite{westerink2008}:

\begin{align}
\frac{\partial \zeta}{\partial t} + \frac{1}{R \cos\phi} \left( \frac{\partial (UH)}{\partial \lambda} + \frac{\partial (VH \cos\phi)}{\partial \phi} \right) &= 0, \\
\frac{\partial U}{\partial t} + \frac{1}{R \cos\phi} U \frac{\partial U}{\partial \lambda} + \frac{V}{R} \frac{\partial U}{\partial \phi} - \left( \frac{\tan\phi}{R} U + f \right) V &= - \frac{1}{R \cos\phi} \frac{\partial}{\partial \lambda} \left[ \frac{P_s}{\rho_0} + g(\zeta - \alpha \eta) \right] \nonumber \\
&\quad + \frac{\nu_T}{H} \frac{\partial}{\partial \lambda} \left( \frac{\partial UH}{\partial \lambda} + \frac{\partial UH}{\partial \phi} \right) + \frac{\tau_{s\lambda}}{\rho_0 H} - \tau_* U, \\
\frac{\partial V}{\partial t} + \frac{1}{R \cos\phi} U \frac{\partial V}{\partial \lambda} + \frac{V}{R} \frac{\partial V}{\partial \phi} + \left( \frac{\tan\phi}{R} V + f \right) U &= - \frac{1}{R} \frac{\partial}{\partial \phi} \left[ \frac{P_s}{\rho_0} + g(\zeta - \alpha \eta) \right] \nonumber \\
&\quad + \frac{\nu_T}{H} \frac{\partial}{\partial \phi} \left( \frac{\partial VH}{\partial \lambda} + \frac{\partial VH}{\partial \phi} \right) + \frac{\tau_{s\phi}}{\rho_0 H} - \tau_* V.
\end{align}

Here, \( \zeta \) is the free-surface elevation, \( U \) and \( V \) are depth-averaged horizontal velocities in the longitudinal and latitudinal directions respectively, \( H = \zeta + h \) is the total water depth (with \( h \) being bathymetry), \( \phi \) and \( \lambda \) denote latitude and longitude, \( f \) is the Coriolis parameter, \( \tau_{s\lambda} \) and \( \tau_{s\phi} \) are wind stress terms, and \( \tau_* \) is the bottom friction term. The model parameters \( \rho_0 \), \( P_s \), \( \alpha \), \( \eta \), \( \nu_T \), and others capture physical effects such as atmospheric pressure, elasticity of the ocean surface, and turbulent mixing.

To support regional surge modeling, the Federal Emergency Management Agency (FEMA) commissioned the development of a synthetic storm surge dataset over the Texas Coast \cite{fema2021}. This dataset comprises 446 tropical cyclone simulations using ADCIRC. Each storm was simulated over a high-resolution unstructured mesh consisting of 3,352,598 nodes and 6,675,517 triangular elements, covering the entire portion of the western North Atlantic Ocean, the Caribbean Sea, and the Gulf west of the 60$^\circ$ meridian. The mesh has exceptionally fine spatial resolution near the Texas and Louisiana coastline (on the order of 10s of meters) and coarser resolution offshore (on the order of a few kilometers). The simulations span a 5-day period with 60 output time steps at 2-hour intervals. Key variables include water elevation, current velocity, wind velocity, surface pressure, and bathymetry. We summarize key dataset properties for a single storm in Table~\ref{tab:table1}.

\begin{table}[h]
\centering
\caption{Summary of FEMA Dataset Key Variables for a Single Storm.}
\begin{tabular}{@{}cccccc@{}}
\toprule
\textbf{ADCIRC File} & \textbf{Variable(s)} & \textbf{Dimensions} & \textbf{Temporal Resolution} & \textbf{Description} & \textbf{Units} \\
\midrule
\texttt{fort.63.nc} & zeta & 60 × 3,352,598 & 2 hours & Water elevation & meters (m) \\
                    & depth & 3,352,598 & N/A & Bathymetry & meters (m) \\
\texttt{fort.64.nc} & u-vel, v-vel & 60 × 3,352,598 & 2 hours & Current velocity & meters/second (m/s) \\
\texttt{fort.73.nc} & pressure & 60 × 3,352,598 & 2 hours & Surface pressure & pascals (Pa) \\
\texttt{fort.74.nc} & windx, windy & 60 × 3,352,598 & 2 hours & Wind velocity & meters/second (m/s) \\
\bottomrule
\end{tabular}
\label{tab:table1}
\end{table}

\subsection{Data Processing}

\begin{figure}[h]
    \centering
    \includegraphics[width=\textwidth]{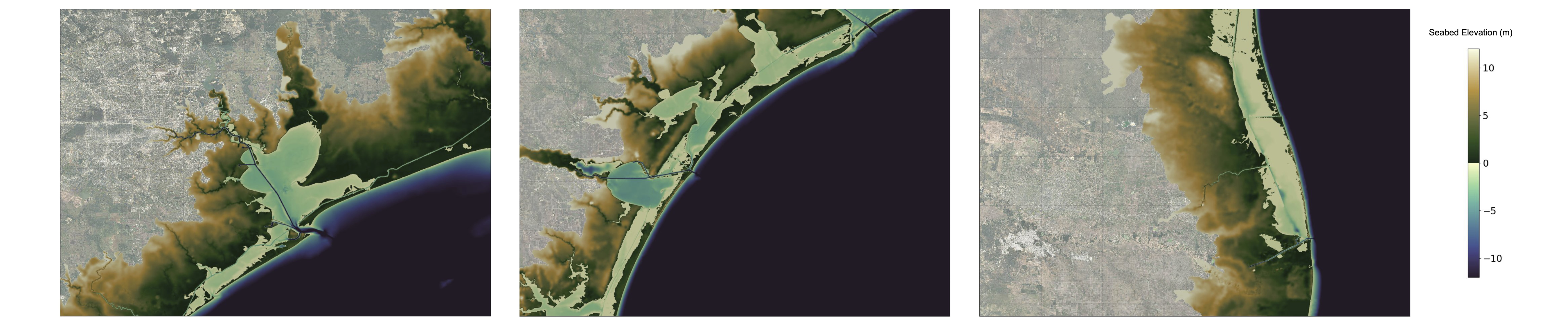}
    \caption{Bathymetric structure of the three selected ROIs: Galveston Bay, Corpus Christi, and South Padre Island.}
    \label{fig:bathy}
\end{figure}

We defined three initial regions of interest (ROIs) along the Texas coastline: Galveston Bay, Corpus Christi Bay, and South Padre Island—ranging from north to south. These regions were selected to capture the diverse physical and geographic characteristics of the Texas coast as visualized in the topo-bathymetry maps shown in Figure~\ref{fig:bathy}. Galveston Bay features shallow, concave bathymetry and complex estuarine dynamics. Corpus Christi lies mid-coast with a semi-enclosed coastal shelf, while South Padre Island opens more directly to the Gulf with deeper nearshore waters. This variation in water depth, coastline geometry, and wind exposure supports learning a robust representation of storm surge behavior and facilitates model generalization across diverse shoreline and forcing regimes.

\begin{figure}[h]
    \centering
    \includegraphics[width=\textwidth]{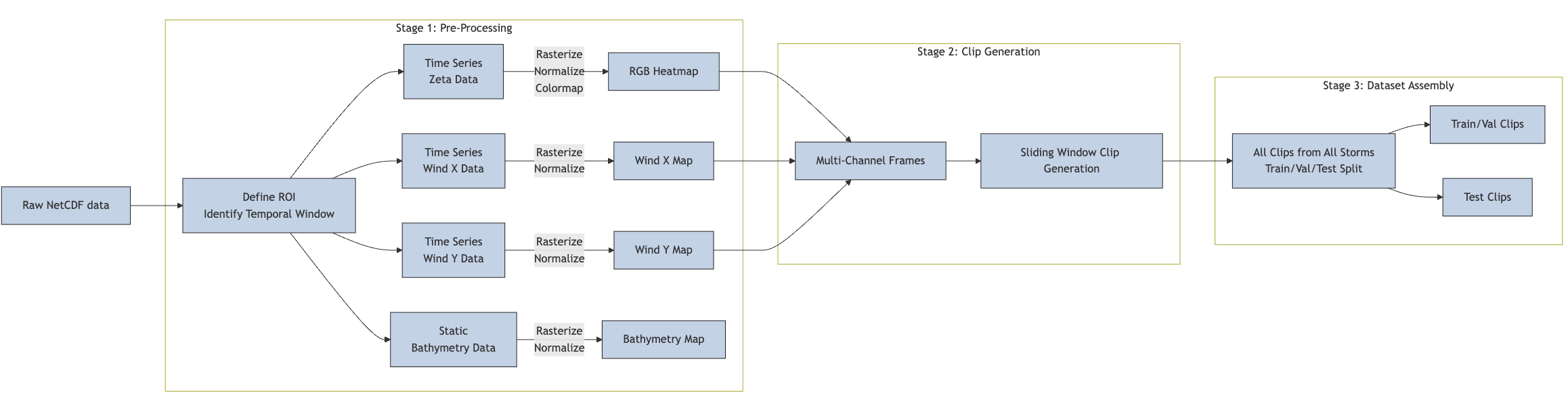}
    \caption{Full processing pipeline from raw ADCIRC NetCDF files to final dataset.}
    \label{fig:pipeline}
\end{figure}

To generate training examples, we projected key ADCIRC output variables—free-surface water elevation (\texttt{zeta}), wind velocity components (\texttt{windx}, \texttt{windy}), and static bathymetry (\texttt{depth})—onto a uniform \(256 \times 256\) grid corresponding to the defined spatial bounding box of each ROI. We chose a \(256 \times 256\) resolution as a balance between computational tractability and spatial fidelity, enabling the model to capture key coastal features while remaining lightweight enough for efficient training.  For example, a \(128 \times 128\) grid exhibited aliasing at key coastal features while a \(512 \times 512\) grid required an inordinate amount of video RAM (VRAM). Rasterization was performed using a triangular mesh projection to preserve spatial fidelity. All variables were mapped based on the empirical distributions observed in the raw data to suppress outliers and improve training stability. Specifically, zeta values were mapped to the range 0–2.5 meters and encoded into RGB heatmaps using a fixed colormap for model input. The upper bound of 2.5 meters was selected based on the 80th percentile of the full dataset’s surge distribution. Wind velocity components were mapped to –40 to 20 m/s in the x-direction and –30 to 30 m/s in the y-direction, with these thresholds chosen to fully encompass the observed range of wind forcing in the dataset. Bathymetry values were mapped between –20 and 50 meters to retain key coastal and nearshore bathymetric features while excluding extreme values that fall outside of the physically meaningful range for shallow water dynamics. The overall data processing workflow is illustrated in Figure~\ref{fig:pipeline}.

After mapping, all variables were linearly scaled to the [0, 1] range. This normalization ensured balanced contributions across modalities and supported stable convergence in convolutional LSTM architectures, which are sensitive to scale disparities between input channels.

\begin{figure}[h]
    \centering
    \includegraphics[width=0.9\textwidth]{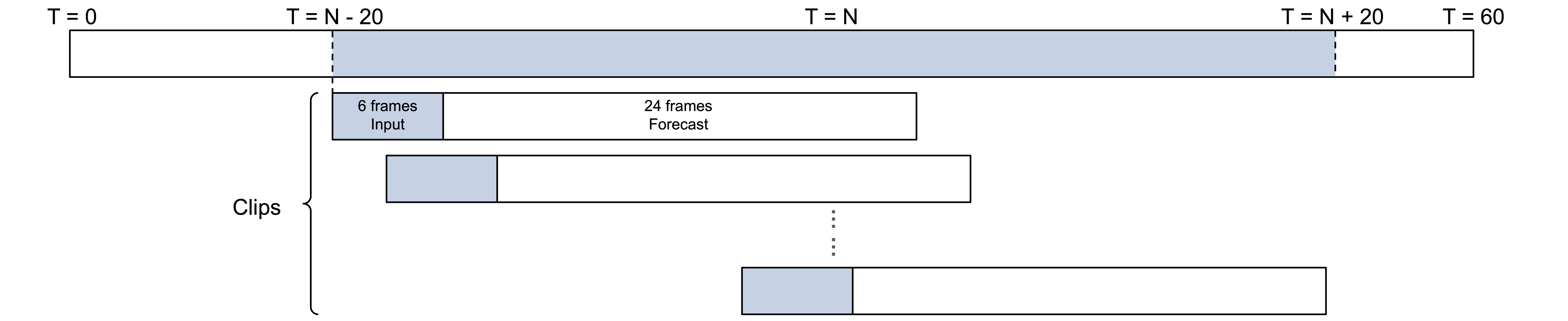}
    \caption{Sliding window clip generation strategy. We identify the peak surge time \( T = N \), then extract a temporal window centered at \( T \). A sliding window of 30 frames is applied to generate input-output samples for training.}
    \label{fig:sliding_window}
\end{figure}

Once the rasterized and normalized variables were generated, we constructed training samples using a fixed-length sliding window of 30 frames centered on the period of peak surge activity. For each storm, \( N \) was identified as the frame at which the average water elevation across all sampled nodes reaches its maximum. A total window spanning from \( N - 20 \) to \( N + 20 \) was extracted to capture the complete temporal evolution of the event, including the buildup, peak, and recession of surge. If the calculated window extended beyond the available time bounds (e.g., near the beginning or end of the simulation), it was truncated accordingly. Each sample consisted of 6 input past frames followed by 24 target frames used for forecasting. The sliding window sampling strategy is illustrated in Figure~\ref{fig:sliding_window}.

Finally, to ensure temporal independence and prevent data leakage, we performed the train/validation/test split at the storm level rather than across individual clips. This guaranteed that all clips originating from a given storm are allocated to the same data partition. By ensuring that the model never encountered data from the same storm during both training and evaluation, we eliminated the possibility of information leakage through temporal correlations. We randomly assigned 90\% of the storms to the training and validation set and reserved the remaining 10\% for testing. 

\begin{figure}[h]
    \centering

    \begin{minipage}{\textwidth}
        \centering
        {\small\text{Galveston Bay}} \\
        \includegraphics[width=\textwidth]{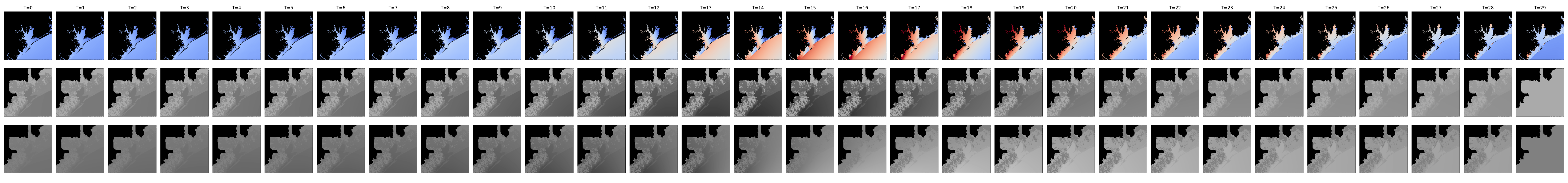}
    \end{minipage}\\[1mm]

    \begin{minipage}{\textwidth}
        \centering
        {\small\text{Corpus Christi}} \\
        \includegraphics[width=\textwidth]{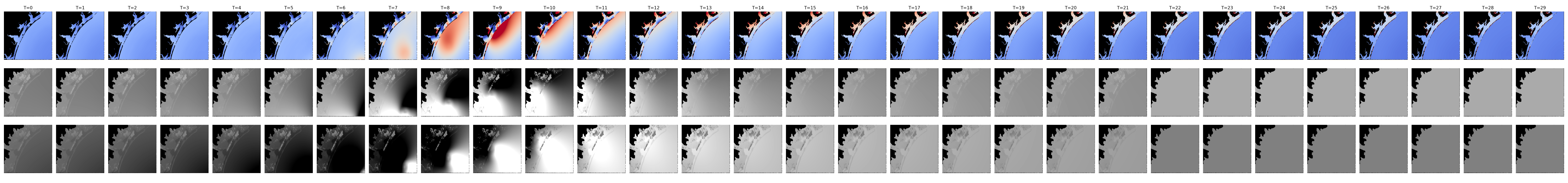}
    \end{minipage}\\[1mm]

    \begin{minipage}{\textwidth}
        \centering
        {\small\text{South Padre Island}} \\
        \includegraphics[width=\textwidth]{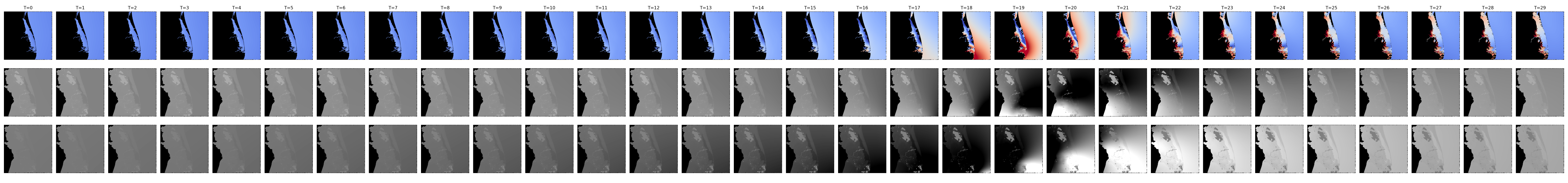}
    \end{minipage}

    \caption{Example clip visualizations from three Texas coastal regions. Each example clip consists of three rows representing different modalities: RGB-encoded water elevation, grayscale x-direction wind velocity, and grayscale y-direction wind velocity. Columns correspond to sequential time frames. These examples highlight the diversity in spatiotemporal surge patterns and atmospheric forcing across different coastal settings.
}
    \label{fig:clip_examples}
\end{figure}

Across the three ROIs, we generated a total of 8,433 training/validation clips and 1,034 testing clips, partitioned as follows: 2,813 train/val and 335 test clips for Galveston Bay, 2,794 train/val and 336 test clips for Corpus Christi, and 2,826 train/val and 363 test clips for South Padre Island. Example training instances produced through this procedure are illustrated in Figure~\ref{fig:clip_examples}.

\section{Methodology}
\label{sec:method}

\subsection{Convolution and Convolutional Neural Networks}

Convolutional neural networks (CNNs) \cite{lecun1998} are a foundational architecture for modeling spatially structured data such as images, videos, and geospatial grids. Their core operation, the convolution, extracts local features by applying learnable filters over spatially coherent neighborhoods.

Given an input tensor \( \mathbf{X} \in \mathbb{R}^{C_{\text{in}} \times H \times W} \) and a convolutional kernel \( \mathbf{K} \in \mathbb{R}^{C_{\text{out}} \times C_{\text{in}} \times k_h \times k_w} \), the convolution operation produces an output tensor \( \mathbf{Y} \in \mathbb{R}^{C_{\text{out}} \times H' \times W'} \) defined as:
\begin{equation}
    \mathbf{Y}_{c, i, j} = \sum_{c'=1}^{C_{\text{in}}} \sum_{m=1}^{k_h} \sum_{n=1}^{k_w} \mathbf{K}_{c, c', m, n} \cdot \mathbf{X}_{c', i + m - p_h, j + n - p_w}
\end{equation}
where \( (p_h, p_w) \) denote the height and width of the padding applied. The padding terms \( (p_h, p_w) \) are typically chosen so that the output maintains the same spatial resolution as the input, i.e., \( H' = H \) and \( W' = W \). Each output channel \( c \) captures a specific learned feature extracted spatially from the input. 

CNNs construct hierarchical feature representations by stacking multiple convolutional layers, interleaved with non-linear activation functions (e.g., ReLU), downsampling operations (e.g., max pooling), and optional normalization or dropout regularization. These networks are particularly effective at capturing translationally invariant patterns, making them suitable for learning complex spatial structures such as coastlines, wind patterns, or elevation gradients.

However, standard CNNs operate only on spatial dimensions and do not account for temporal evolution, limiting their direct applicability to forecasting tasks.

\subsection{Long Short-Term Memory}

Recurrent neural networks (RNNs) \cite{rumelhart1986} are a natural choice for modeling sequential data, owing to their recursive structure, which allows information to propagate across time. However, standard RNNs suffer from vanishing or exploding gradient problems \cite{bengio1994}, which hinder their ability to capture long-term dependencies. To address this shortcoming, Hochreiter and Schmidhuber in~\cite{hochreiter1997}, introduced the Long Short-Term Memory (LSTM) architecture.

An LSTM cell is designed to maintain an internal memory vector \( \mathbf{c}_t \) that can be selectively updated, erased, or read via gating mechanisms. At each time step \( t \), given an input vector \( \mathbf{x}_t \in \mathbb{R}^d \) and a hidden state from the previous time step \( \mathbf{h}_{t-1} \in \mathbb{R}^h \), the cell performs the following computations:

First, the forget gate \( \mathbf{f}_t \in [0,1]^h \) determines which parts of the previous memory \( \mathbf{c}_{t-1} \) to retain:

\begin{equation} \label{eq:forgetgate}
    \mathbf{f}_t = \sigma\left( \mathbf{W}_f [\mathbf{h}_{t-1}, \mathbf{x}_t] + \mathbf{b}_f \right).
\end{equation}

Next, the input gate \( \mathbf{i}_t \in [0,1]^h \) regulates the incorporation of new information, and a candidate vector \( \tilde{\mathbf{c}}_t \in \mathbb{R}^h \) is computed:

\begin{align} \label{eq:inputgate}
    \mathbf{i}_t &= \sigma\left( \mathbf{W}_i [\mathbf{h}_{t-1}, \mathbf{x}_t] + \mathbf{b}_i \right), \\
    \tilde{\mathbf{c}}_t &= \tanh\left( \mathbf{W}_c [\mathbf{h}_{t-1}, \mathbf{x}_t] + \mathbf{b}_c \right).
\end{align}

The cell state is then updated by combining the previous memory and new candidate information:

\begin{equation} \label{eq:cell_state}
    \mathbf{c}_t = \mathbf{f}_t \odot \mathbf{c}_{t-1} + \mathbf{i}_t \odot \tilde{\mathbf{c}}_t.
\end{equation}

Finally, the output gate \( \mathbf{o}_t \in [0,1]^h \) determines which parts of the updated memory to expose through the hidden state:

\begin{align} \label{eq:outgate}
    \mathbf{o}_t &= \sigma\left( \mathbf{W}_o [\mathbf{h}_{t-1}, \mathbf{x}_t] + \mathbf{b}_o \right), \\
    \mathbf{h}_t &= \mathbf{o}_t \odot \tanh(\mathbf{c}_t).
\end{align}

Here, the sigmoid and hyperbolic tangent activation functions are defined as follows:

\begin{equation} \label{eq:activation_func}
    \sigma(x) = \frac{1}{1 + e^{-x}}, \qquad \tanh(x) = \frac{e^x - e^{-x}}{e^x + e^{-x}},
\end{equation}

The operator \( \odot \) denotes element-wise multiplication. The trainable parameters in equations~\eqref{eq:forgetgate} through~\eqref{eq:outgate} include the weight matrices \( (\mathbf{W}_f, \mathbf{W}_i, \mathbf{W}_c, \mathbf{W}_o) \) and the biases \( (\mathbf{b}_f, \mathbf{b}_i, \mathbf{b}_c, \mathbf{b}_o) \).

This gating structure enables LSTM networks to learn both short and long range temporal dependencies while mitigating the vanishing gradient problem, making them a foundational architecture for sequence modeling.

\subsection{Convolutional LSTM for Spatiotemporal Forecasting}

\subsubsection{Model Architecture}
While LSTM networks are highly effective in capturing short and long range temporal dependencies, they operate on flattened one-dimensional sequences and are therefore agnostic to spatial correlations across structured inputs such as images or geophysical fields. For example, applying a standard LSTM to a sequence of 2D water elevation maps would require reshaping each frame into a vector, thereby discarding spatial locality, which is a critical aspect of storm surge dynamics. To overcome this limitation, Shi \emph{et al.} in \cite{shi2015} proposed the Convolutional LSTM (ConvLSTM) architecture, which replaces the fully connected transformations in the gating mechanisms with convolutional operations. This modification preserves spatial structure while maintaining the temporal memory capability of LSTMs, enabling the model to jointly learn spatiotemporal patterns from gridded data.

Let \( \mathbf{X}_t \in \mathbb{R}^{C \times H \times W} \) denote the input tensor at time \( t \), where \( C \) is the number of input channels and \( H \times W \) is the spatial resolution. Let \( D \) denote the number of hidden channels used to represent the internal memory of the model. Given the previous hidden state \( \mathbf{H}_{t-1} \in \mathbb{R}^{D \times H \times W} \) and cell state \( \mathbf{C}_{t-1} \in \mathbb{R}^{D \times H \times W} \), the ConvLSTM, similar to the standard LSTM, maintains and updates a memory cell using four convolutionally computed gates.

The input and hidden states are first concatenated along the channel dimension and passed through a shared convolutional kernel to produce the forget, input, candidate, and output gates. Each gate preserves spatial structure and is applied element-wise across the field.

The forget gate \( \mathbf{F}_t \in [0,1]^{D \times H \times W} \) controls how much of the previous memory to retain:

\begin{equation} \label{eq:conv_forget}
    \mathbf{F}_t = \sigma\left( \mathbf{W}_f * [\mathbf{X}_t, \mathbf{H}_{t-1}] + \mathbf{b}_f \right).
\end{equation}

The input gate \( \mathbf{I}_t \) determines which parts of the new candidate state to incorporate, and the candidate cell state \( \tilde{\mathbf{C}}_t \) is computed using a \(\tanh\) activation:

\begin{align} \label{eq:colv_input}
    \mathbf{I}_t &= \sigma\left( \mathbf{W}_i * [\mathbf{X}_t, \mathbf{H}_{t-1}] + \mathbf{b}_i \right), \\
    \tilde{\mathbf{C}}_t &= \tanh\left( \mathbf{W}_c * [\mathbf{X}_t, \mathbf{H}_{t-1}] + \mathbf{b}_c \right).
\end{align}

The new cell state \( \mathbf{C}_t \) is then formed by combining the retained memory and the gated candidate:

\begin{equation} \label{eq:conv_new_state}
    \mathbf{C}_t = \mathbf{F}_t \odot \mathbf{C}_{t-1} + \mathbf{I}_t \odot \tilde{\mathbf{C}}_t
\end{equation}

The output gate \( \mathbf{O}_t \in [0,1]^{D \times H \times W} \) determines how much of the internal state to expose:

\begin{align} \label{eq:conv_outgate}
    \mathbf{O}_t &= \sigma\left( \mathbf{W}_o * [\mathbf{X}_t, \mathbf{H}_{t-1}] + \mathbf{b}_o \right) \\
    \mathbf{H}_t &= \mathbf{O}_t \odot \tanh(\mathbf{C}_t)
\end{align}

Here, \( * \) denotes a 2D convolution, and \( \sigma \) and \( \tanh \) refer to the sigmoid and hyperbolic tangent activation functions, respectively, as defined in equation~\eqref{eq:activation_func}. The parameters \( (\mathbf{W}_f, \mathbf{W}_i, \mathbf{W}_c, \mathbf{W}_o) \) and the corresponding biases \( (\mathbf{b}_f, \mathbf{b}_i, \mathbf{b}_c, \mathbf{b}_o) \) are learned convolutional kernels shared across spatial locations.

In our model, each input frame consisted of six channels: three for RGB-encoded water elevation, two for wind velocity components (x and y), and one for static bathymetry. The model was trained to predict the future evolution of the RGB water elevation field over 24 frames, conditioned on a 6-frame historical context.

\begin{figure}[h]
    \centering
    \includegraphics[width=\textwidth]{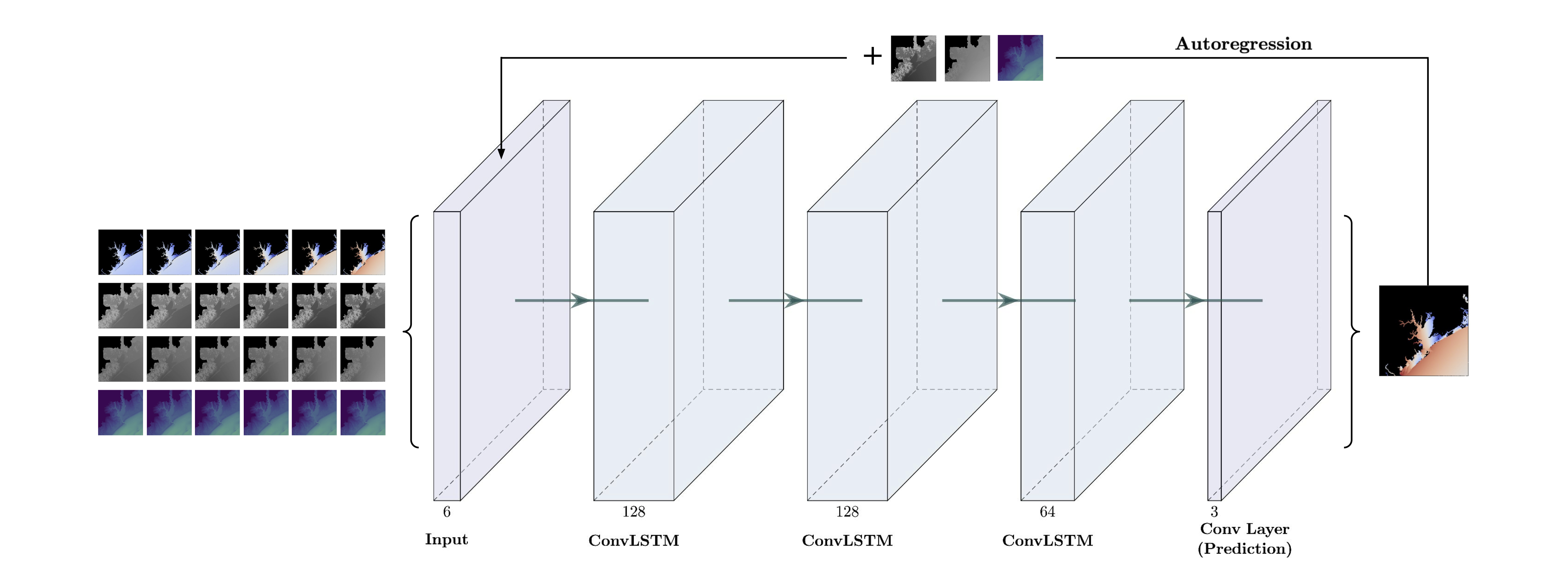}
    \caption{Model architecture consisting of three stacked ConvLSTM layers followed by a $1 \times 1$ convolutional decoder. The input consists of RGB-encoded water elevation, wind velocity components, and static bathymetry. Forecasting is conducted autoregressively, where at each future time step, the predicted RGB frame is concatenated with the corresponding wind fields and static bathymetry to form the input for the next step. The model predicts future RGB water elevation fields over 48 hours.}

    \label{fig:model}
\end{figure}

The network architecture consisted of three stacked ConvLSTM layers, with hidden channel dimensions \( D = [128, 128, 64] \), each employing a \( 3 \times 3 \) convolutional kernel with padding to preserve spatial resolution. The final output from the topmost ConvLSTM layer was passed through a \( 1 \times 1 \) convolutional decoder to produce a 3-channel RGB prediction. This configuration is illustrated in Figure~\ref{fig:model}.

\subsubsection{Forecasting Strategy}
Forecasting was conducted in an autoregressive manner: at each future time step \( t \), the model produced a predicted RGB field \( \hat{\mathbf{Y}}_t \), which was used as input in the next step, along with the corresponding ground-truth wind fields and static bathymetry. The effective input to the ConvLSTM at time \( t \) was:

\begin{equation} \label{eq:conLSTM_input}
    \mathbf{X}_t = \left[ \hat{\mathbf{Y}}_{t-1}, \mathbf{W}_{x,t}, \mathbf{W}_{y,t}, \mathbf{B} \right] \in \mathbb{R}^{6 \times H \times W},
\end{equation}

where \( \hat{\mathbf{Y}}_{0} \) was the final RGB frame in the input context window, \( \mathbf{W}_{x,t} \) and \( \mathbf{W}_{y,t} \) were the exogenous wind forcings at time \( t \), and \( \mathbf{B} \) was the spatially fixed bathymetric field. Importantly, wind fields were treated as known future inputs, which is realistic in operational settings where wind forecasts are available from external atmospheric models. Bathymetry, though static, provides essential spatial context: by conditioning on local water depth, the model can infer complex interactions between surge propagation, wind stress, and coastal geometry through convolutional receptive fields.

To promote robustness during training and reduce compounding errors during rollout, teacher forcing was employed, where the ground-truth frame \( \mathbf{Y}_{t-1} \) was probabilistically substituted in place of \( \hat{\mathbf{Y}}_{t-1} \). Additionally, dropout regularization with probability \( p = 0.1 \) was applied between ConvLSTM layers to mitigate overfitting and encourage generalization.

This integration allowed the network to forecast future surge fields not solely based on past surge observations, but also by incorporating physically grounded drivers such as wind and bathymetry. As a result, the model was not constrained to memorizing region-specific surge patterns, but instead learned generalizable spatiotemporal relationships that can support extrapolation to a broad range of coastal settings and storm scenarios.

\section{Configuration and Evaluation}

\label{sec:config}

\subsection{Hardware and Configuration Settings}

All experiments were conducted on the TACC Vista system \cite{vista}, a cutting-edge high-performance computing platform equipped with NVIDIA GH200 Grace Hopper Superchips. For each experiment, we utilized a single node containing one GH200 GPU, which combines a 72-core Grace CPU and a 96 GB Hopper GPU, providing large shared-memory bandwidth ideal for deep learning workloads.

We trained separate models for each ROI: Galveston Bay, Corpus Christi Bay, and South Padre Island, as well as an additional model trained on a combined dataset from all three regions. Across all models, we used a batch size of 3 clips, an initial learning rate of 0.001, a total of 50 training epochs, and a dropout probability of 0.1 applied between ConvLSTM layers. Each ConvLSTM layer used a \(3 \times 3\) kernel, and the hidden state dimensions were fixed at \([128, 128, 64]\). The learning rate was automatically adjusted throughout training using a learning rate scheduler that dynamically reduces the step size in response to stagnation in validation loss. These hyperparameters, including model depth, convolutional configuration, and regularization settings, were selected based on a grid search over the validation set to ensure balanced model complexity and generalization. The model was trained using the Adam optimizer and mean squared error loss, computed between predicted and ground-truth RGB water elevation fields at the pixel level.

For the combined-region model, in addition to the standard training setup described above, we adopted a memory-efficient lazy loading technique that avoided loading the entire dataset into memory. Specifically, we distributed the training data across three data files and loaded samples on-the-fly during each call using a custom PyTorch dataset. Each sample included RGB-encoded water elevation, wind velocity components in the \(x\) and \(y\) directions, and a region-specific static bathymetry map, which was preloaded once and cached for fast access.

\subsection{Evaluation Metrics}

We report prediction performance using standard image reconstruction metrics: mean squared error (MSE), mean absolute error (MAE), root mean squared error (RMSE), and coefficient of determination (\( R^2 \)). Let \( \hat{\mathbf{Y}}_{n,t} \in \mathbb{R}^{C \times H \times W} \) be the predicted RGB frame at time \( t \) for test clip \( n \), and \( \mathbf{Y}_{n,t} \) the corresponding ground truth. Let \( T = 24 \) be the number of predicted frames per clip, and \( H = W = 256 \), \( C = 3 \). For each forecast time \( t \), we computed the per-frame metric for each clip by flattening all spatial and channel dimensions:

\begin{equation}
\text{MSE}_{n,t} = \frac{1}{HWC} \| \hat{\mathbf{Y}}_{n,t} - \mathbf{Y}_{n,t} \|_2^2
\end{equation}

\begin{equation}
\text{MAE}_{n,t} = \frac{1}{HWC} \| \hat{\mathbf{Y}}_{n,t} - \mathbf{Y}_{n,t} \|_1
\end{equation}

\begin{equation}
\text{RMSE}_{n,t} = \sqrt{\text{MSE}_{n,t}}
\end{equation}

\begin{equation}
R^2_{n,t} = 1 - \frac{ \| \hat{\mathbf{Y}}_{n,t} - \mathbf{Y}_{n,t} \|_2^2 }{ \| \mathbf{Y}_{n,t} - \bar{\mathbf{Y}}_{n,t} \|_2^2 }
\end{equation}

where \( \bar{\mathbf{Y}}_{n,t} \in \mathbb{R}^{C \times H \times W} \) denotes the mean pixel value averaged over all spatial and channel dimensions of the ground truth frame \( \mathbf{Y}_{n,t} \). For each time step \( t \), we summarized the distribution of \(\text{MSE}_{n,t}\), \(\text{RMSE}_{n,t}\), \(\text{MAE}_{n,t}\), and \(R^2_{n,t}\) across all test clips \( n = 1, \dots, N \) to evaluate frame-wise prediction accuracy over time.

\section{Results and Discussion}
\label{sec:result}

\subsection{Model Accuracy}

In this section, we assess the model's accuracy across different coastal regions using both region-specific and combined-region training strategies. The first part of this section details results from region-specific models, each trained exclusively on a single region of interest (ROI): Galveston Bay, Corpus Christi, or South Padre Island, and evaluated on test clips from the same region. This design allows us to measure the model’s performance in learning region-specific surge dynamics, as detailed in the following subsection. To isolate the contribution of physics-informed auxiliary inputs, we also evaluate an ablation model trained without wind forcing and bathymetry. This baseline allows us to quantify the added value of physical context in the learning process.

The second part of this section details the performance of a model trained on the aggregated training data from all three ROIs. This combined-region model was evaluated on the union of all test clips across ROIs, allowing us to investigate whether broader exposure improves generalization across diverse coastal geometries and forcing conditions. We include a side-by-side comparison of this model’s performance against each region-specific model, highlighting performance trade-offs and potential gains from joint training.

\subsubsection{Regional Performance Across Three Coastal Domains}

\begin{figure}[htbp]
    \centering
    {\small\text{Prediction Visualization for a Single Galveston Bay Example}}\\
    \includegraphics[width=0.7\textwidth]{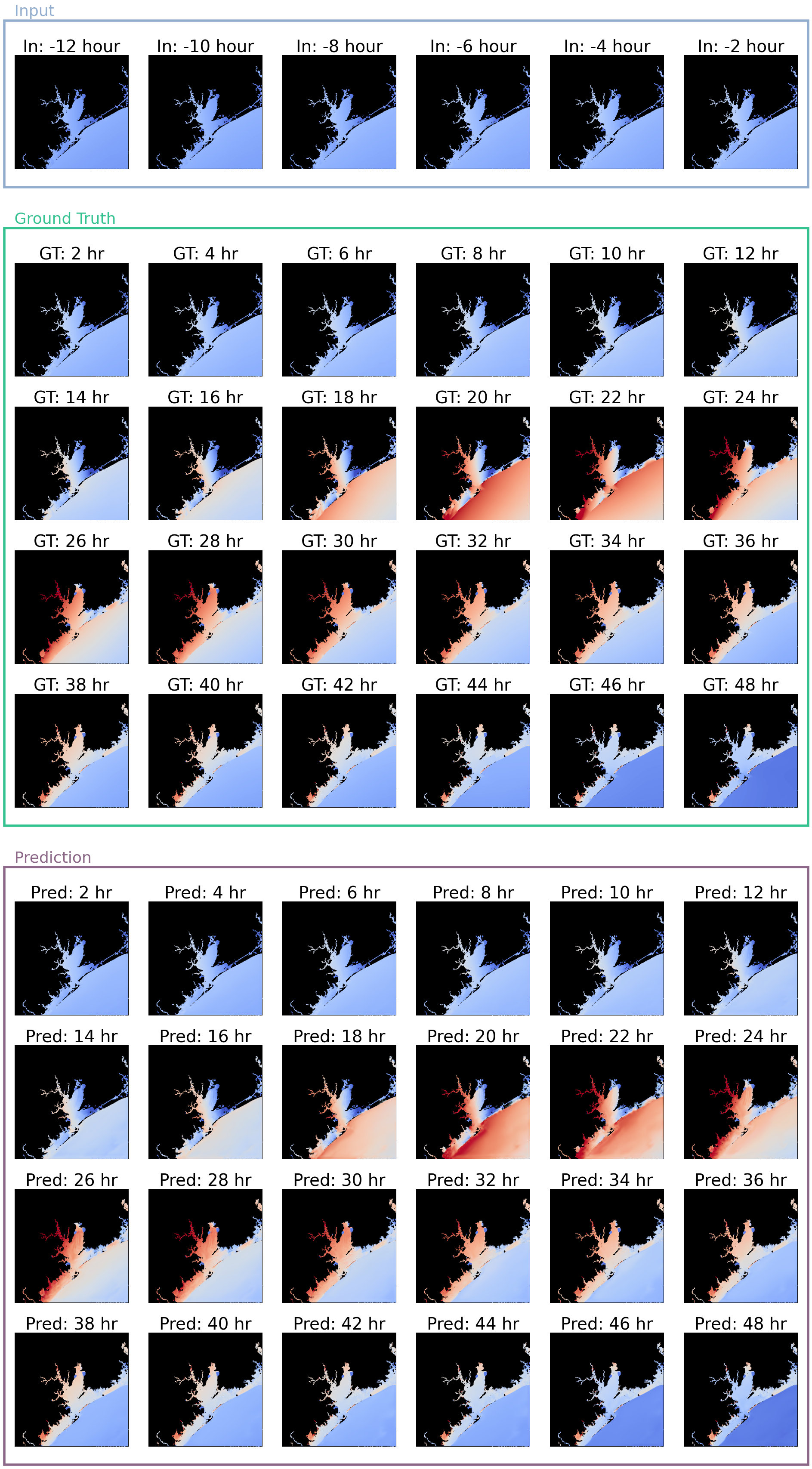}\\[1ex]

    \caption{A single example from Galveston Bay shows predicted water elevation over time. The first box (Input) contains 6 context frames of past water elevation. The second box (Ground Truth) shows the ground truth future evolution, and the third box (Prediction) displays the model's prediction.}

    \label{fig:galveston_example}
\end{figure}

\begin{figure}[htbp]
    \centering
    {\small\text{Prediction Visualization for a Single Corpus Christi Example}}\\
    \includegraphics[width=0.7\textwidth]{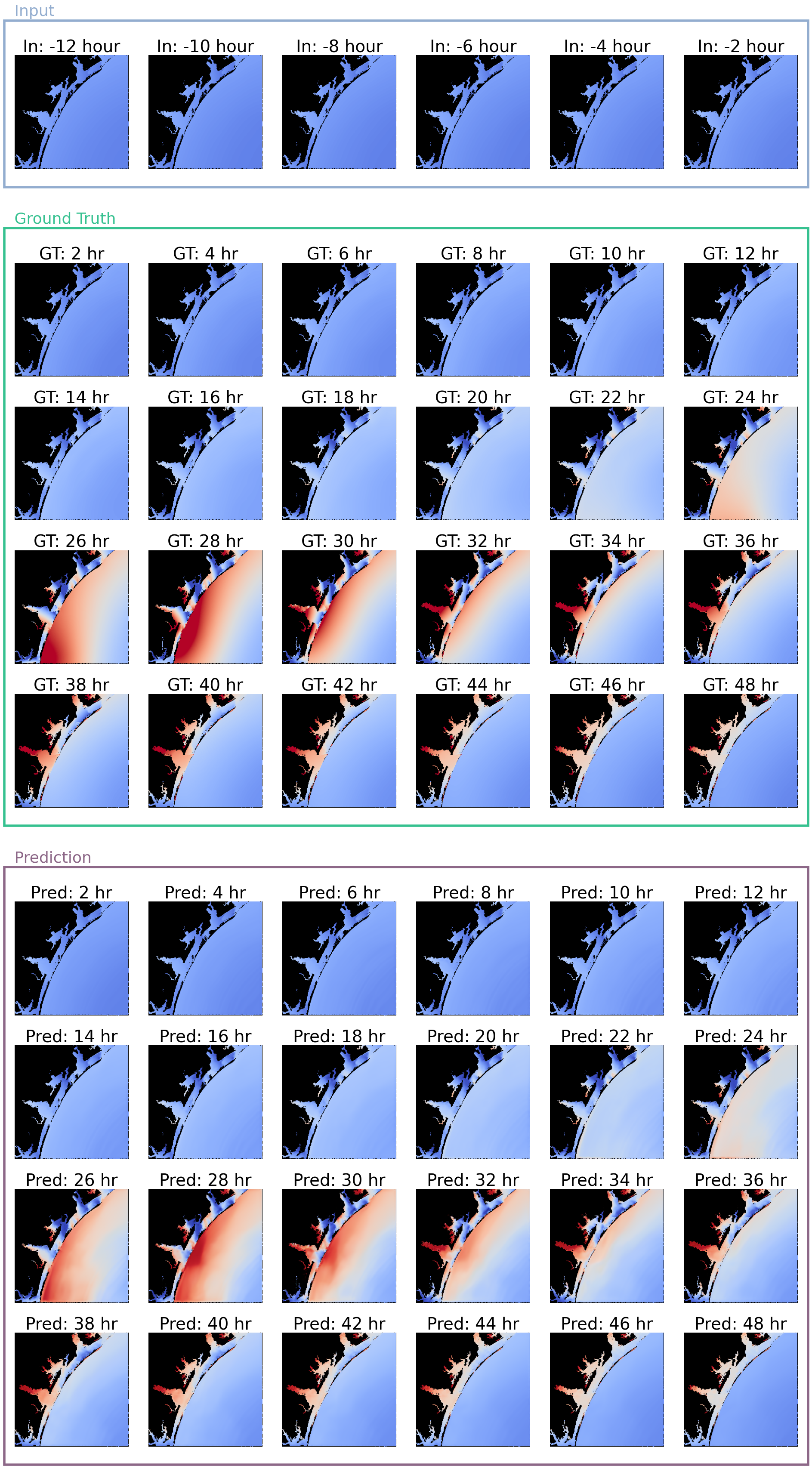}\\[1ex]

    \caption{A single example from Corpus Christi shows predicted water elevation over time. The first box (Input) contains 6 context frames of past water elevation. The second box (Ground Truth) shows the ground truth future evolution, and the third box (Prediction) displays the model's prediction.}

    \label{fig:corpus_example}
\end{figure}

\begin{figure}[htbp]
    \centering
    {\small\text{Prediction Visualization for a Single South Padre Island Example}}\\
    \includegraphics[width=0.7\textwidth]{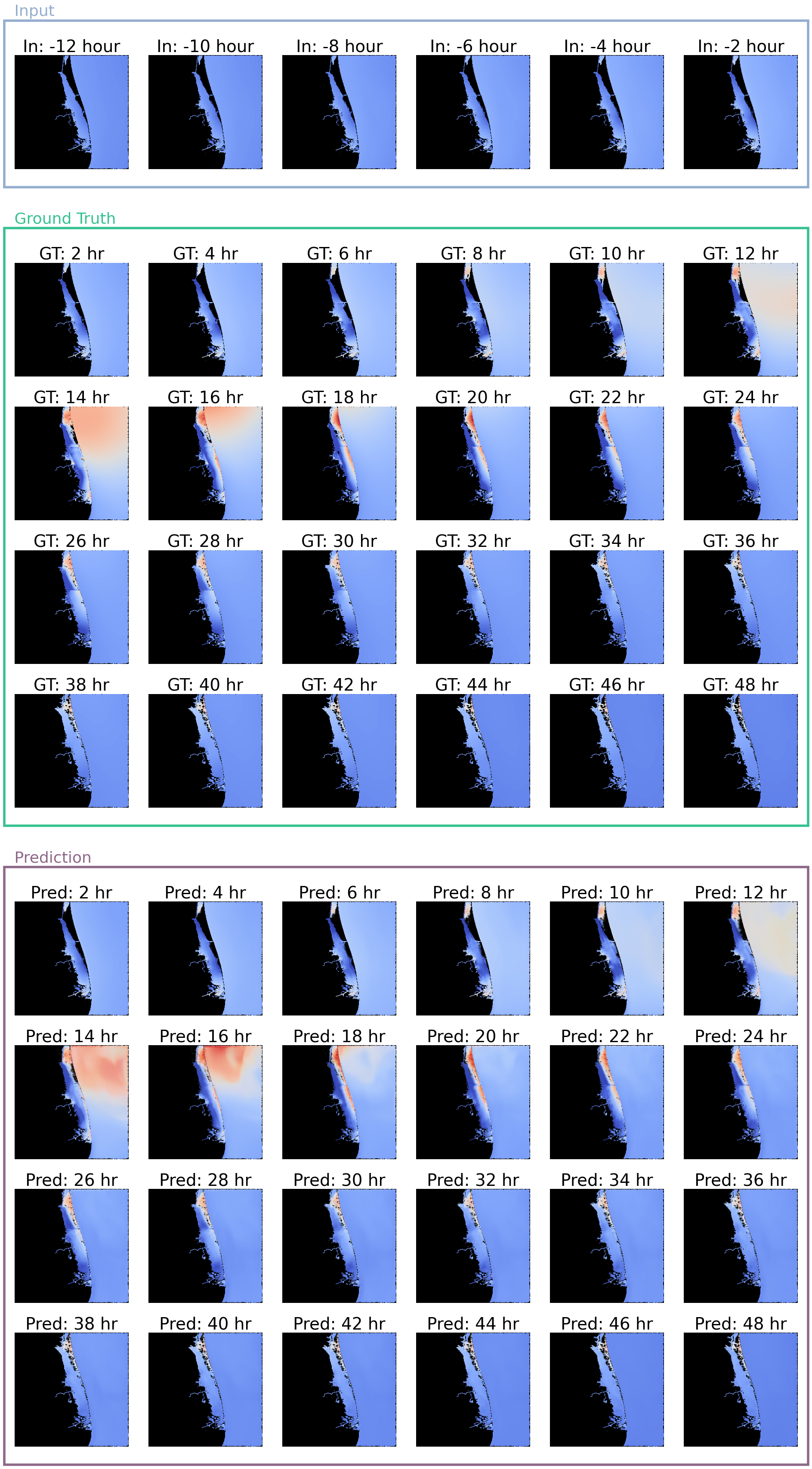}\\[1ex]
    
    \caption{A single example from South Padre Island shows predicted water elevation over time. The first box (Input) contains 6 context frames of past water elevation. The second box (Ground Truth) shows the ground truth future evolution, and the third box (Prediction) displays the model's prediction.}

    \label{fig:south_example}
\end{figure}

For the regional models, we trained and evaluated three independent ConvLSTM models, each tailored to a specific coastal region in Texas. The Galveston Bay model was trained on 2,813 clips and tested on 335 clips; the Corpus Christi model used 2,794 training and 336 test clips; and the South Padre Island model was trained on 2,826 and evaluated on 363 clips. Each clip consisted of 6 contextual RGB input frames, followed by 24 prediction frames, with wind fields and static bathymetry provided as auxiliary input channels. Representative qualitative and quantitative results are shown in Figures~\ref{fig:galveston_example}, \ref{fig:corpus_example}, and~\ref{fig:south_example}, where each box illustrates the contextual input, ground truth surge progression, and predicted output, respectively. This approach to visual, spatially continuous forecasting is critical not only for assessing prediction quality but also for practical utility, as decision makers in emergency response often require interpretable, region-wide surge evolution rather than raw scalar outputs at isolated coordinates.

The predictions demonstrate compelling accuracy across multiple dimensions: surge intensity, timing of coastal inundation onset, and coastline transformation are all closely aligned with the ground truth. One particularly striking capability is the model’s handling of fine-scale coastal hydrodynamics. In all three regions, the model is able to capture intricate surge dynamics within narrow estuaries and river networks. For example, the rise and fall of water levels in Buffalo Bayou near Houston, Oso Bay near Corpus Christi, and the sinuous channels cutting through the barrier zone in South Padre are resolved with remarkable spatial detail. In some cases, the model even differentiates vertical gradients within individual river channels, for example, a higher surge in downstream reaches and lower values upstream, indicating a deep spatiotemporal understanding of fluid transport. Even more impressive is the model’s ability to predict geomorphologically complex outcomes, such as the complete temporary inundation of the barrier island along Padre Island National Seashore. Notably, the model also performs well in low-activity frames, avoiding overfitting to surge-heavy regimes or specific coastal geometries.

\begin{figure}[htbp]
    \centering

    {\small\text{Prediction Accuracy Over Forecast Time for the Same Example}}\\
    \includegraphics[width=\textwidth]{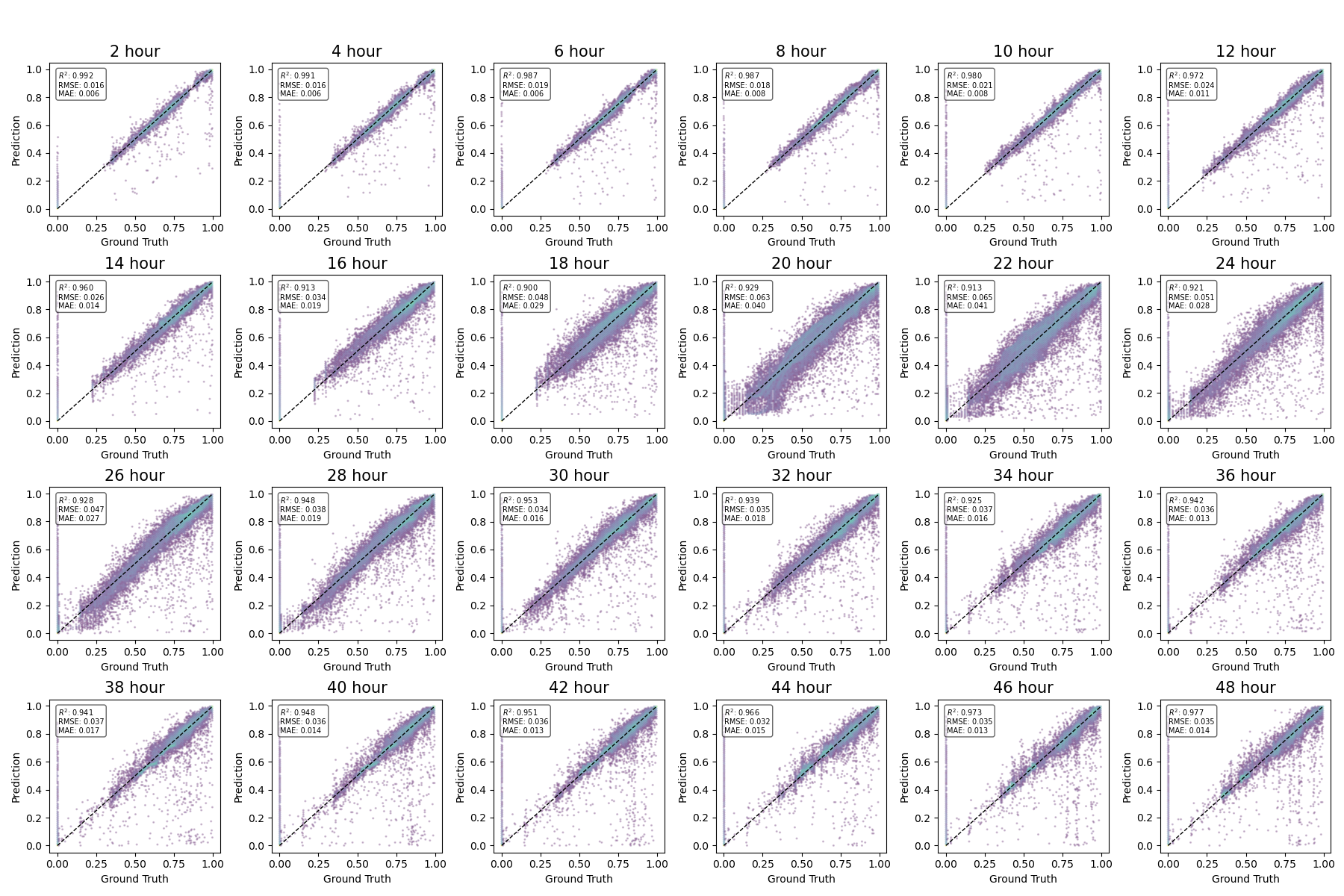}
    
    \caption{Prediction accuracy metrics ($R^2$, RMSE, and MAE) at each forecast hour for the same Galveston Bay example shown in Figure~\ref{fig:galveston_example}.}

    \label{fig:galveston_metrics}
\end{figure}

\begin{figure}[htbp]
    \centering
    
    {\small\text{Prediction Accuracy Over Forecast Time for the Same Example}}\\
    \includegraphics[width=\textwidth]{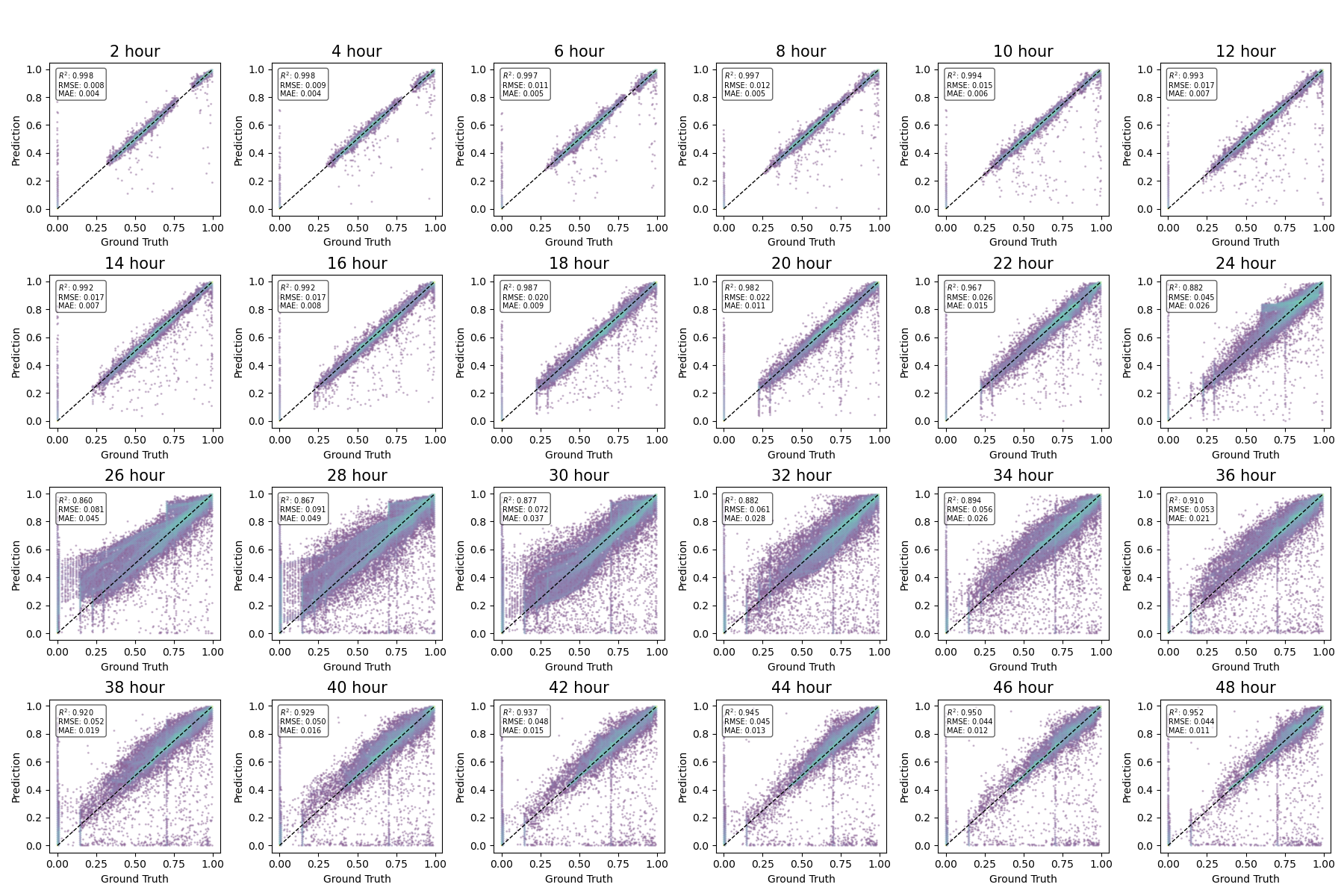}
    
    \caption{Prediction accuracy metrics ($R^2$, RMSE, and MAE) at each forecast hour for the same Corpus Christi example shown in Figure~\ref{fig:corpus_example}.}

    \label{fig:corpus_metrics}
\end{figure}

\begin{figure}[htbp]
    \centering
    
    {\small\text{Prediction Accuracy Over Forecast Time for the Same Example}}\\
    \includegraphics[width=\textwidth]{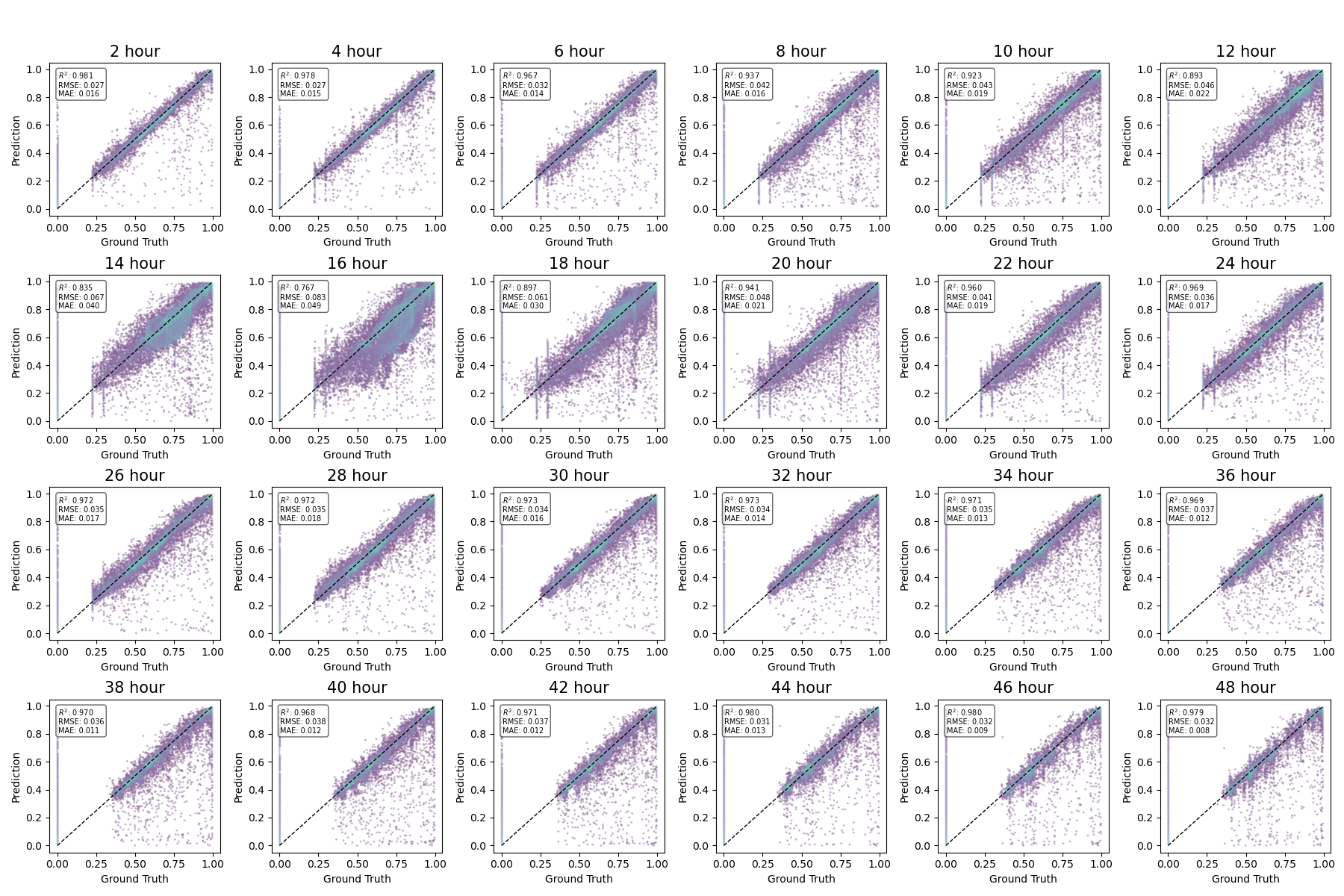}
    
    \caption{Prediction accuracy metrics ($R^2$, RMSE, and MAE) at each forecast hour for the same South Padre Island example shown in Figure~\ref{fig:south_example}.}

    \label{fig:south_metrics}
\end{figure}

Quantitative performance for these representative examples is summarized in the accompanying per-frame \( \text{R}^2 \), RMSE, and MAE plots, aggregated across all RGB pixels, shown in Figures~\ref{fig:galveston_metrics}, \ref{fig:corpus_metrics}, and~\ref{fig:south_metrics}. As expected, peak surge moments show slight degradation, for example, frame 20 for Galveston (\( \text{R}^2 = 0.900 \)), frame 20 for Corpus Christi (\( \text{R}^2 = 0.860 \)), and frame 16 for South Padre (\( \text{R}^2 = 0.767 \)), but overall prediction skill remains strong. A consistent pattern of underprediction is observed across all three examples. This behavior aligns with prior observations that ConvLSTM and other convolutional RNNs trained with MSE loss often yield smoothed or blurred predictions, especially in high-variance or multimodal regions, thereby underestimating sharp surge peaks \cite{shi2015, mathieu2016}. Our use of pre-smoothed RGB heatmap data helps mitigate this effect by preserving spatial detail and enhancing visual coherence.

\begin{figure}[htbp]
    \centering
    {\small\text{Galveston Bay - Regional Performance Overview}}\\
    \includegraphics[width=\textwidth]{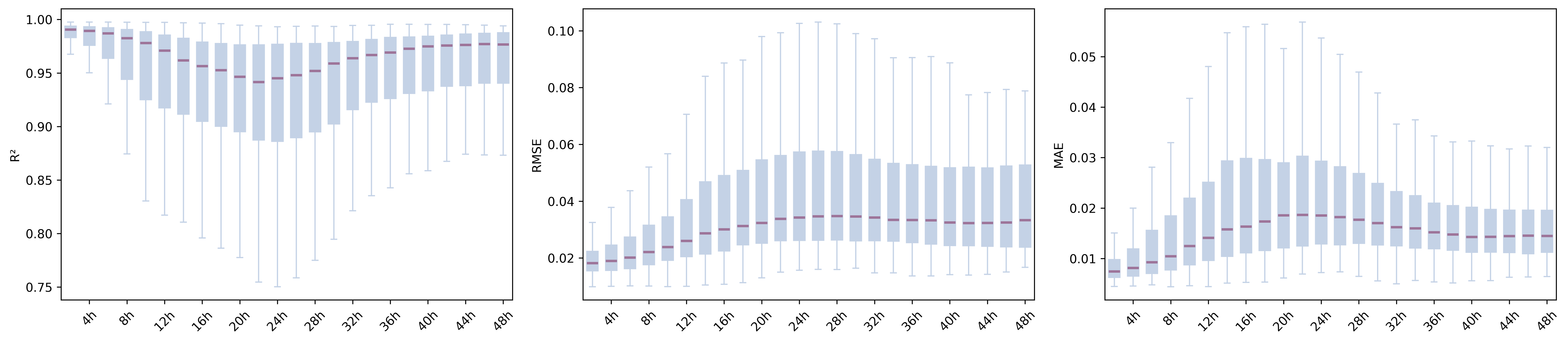}
    
    \caption{Prediction accuracy across all test clips in the Galveston Bay region. Each box represents the distribution of per-frame $R^2$, RMSE, and MAE scores computed across all clips at each forecast hour. Compared to the single-clip result, this visualization summarizes the model's generalization performance across the entire region.}
    
    \label{fig:galveston_general}
\end{figure}

\begin{figure}[htbp]
    \centering
    {\small\text{Corpus Christi - Regional Performance Overview}}\\
    \includegraphics[width=\textwidth]{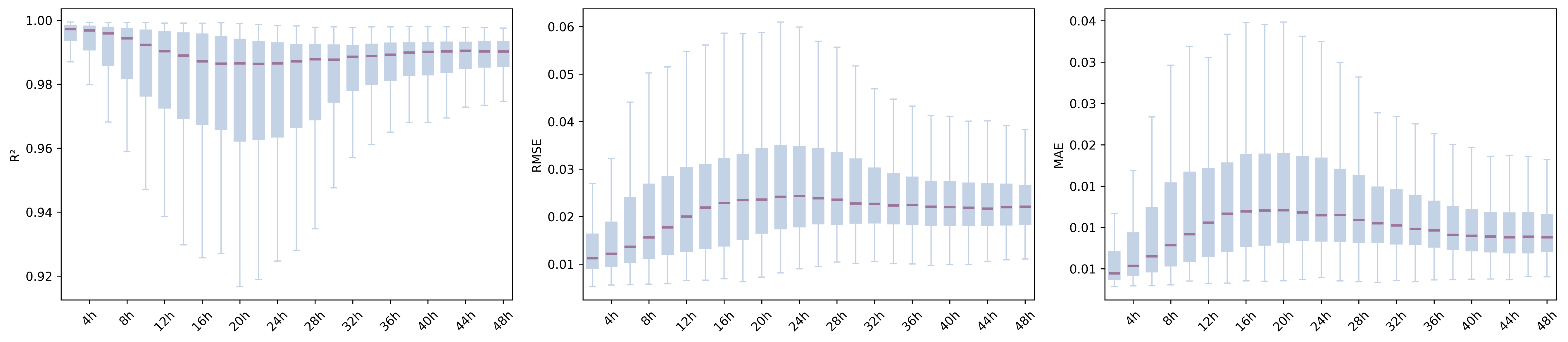}
    
    \caption{Prediction accuracy across all test clips in the Corpus Christi region. Each box represents the distribution of per-frame $R^2$, RMSE, and MAE scores computed across all clips at each forecast hour. Compared to the single-clip result, this visualization summarizes the model's generalization performance across the entire region.}
    
    \label{fig:corpus_general}
\end{figure}

\begin{figure}[htbp]
    \centering
    {\small\text{South Padre Island - Regional Performance Overview}}\\
    \includegraphics[width=\textwidth]{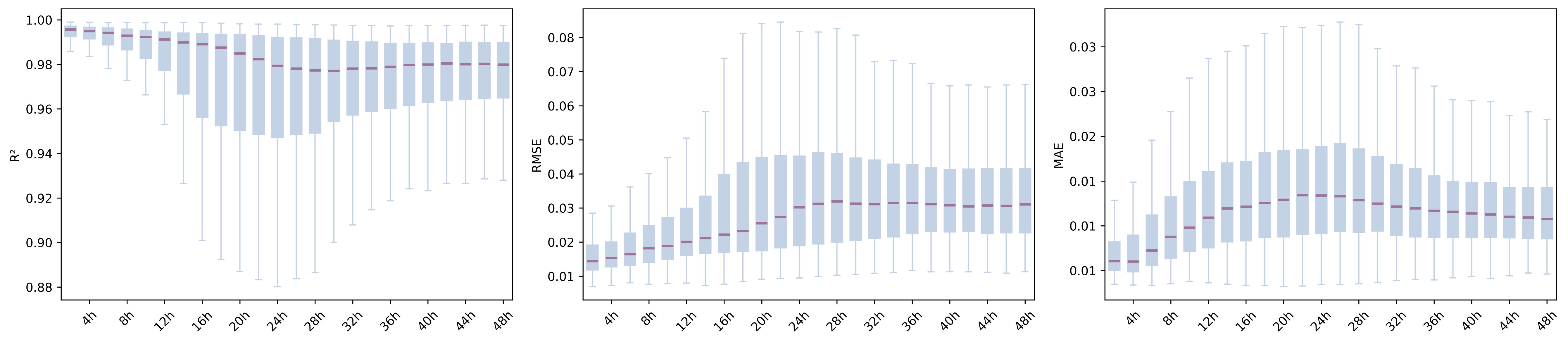}
    
    \caption{Prediction accuracy across all test clips in the South Padre Island region. Each box represents the distribution of per-frame $R^2$, RMSE, and MAE scores computed across all clips at each forecast hour. Compared to the single-clip result, this visualization summarizes the model's generalization performance across the entire region.}
    
    \label{fig:south_general}
\end{figure}

We also present regional generalization results in Figures~\ref{fig:galveston_general}, \ref{fig:corpus_general}, and~\ref{fig:south_general}, with corresponding summary statistics reported in Tables~\ref{tab:median_r2}, \ref{tab:median_rmse}, and~\ref{tab:median_mae}. Each figure displays the distribution of per-frame \(R^2\), RMSE, and MAE scores across all test clips in the respective region. Specifically, at each forecast hour, we computed the metrics measured from predicted and ground truth RGB frames for every test clip, then visualized the distribution of these values as a boxplot. The solid purple line indicates the median; the box spans the interquartile range (IQR) from the first quartile (Q1) to the third quartile (Q3); and the whiskers extend to 1.5 times the IQR or to the most extreme data points within that range.

Across all three regions, a consistent temporal trend emerges: model accuracy degrades slightly around forecast hour 24, which often aligns with periods of heightened surge activity. This behavior reflects the way training and test clips were constructed, each clip spans a ±20-frame window centered around the peak surge frame of a given storm. Consequently, the most dynamic and nonlinear behavior, typically involving rapid spatial transitions in water elevation, tends to cluster near the middle of the forecast horizon. These frames pose greater predictive challenges due to their complexity. While the exact timing of peak dynamics varies across events, this temporal concentration of surge activity is an inherent characteristic of the dataset design.

Among the three regions, Galveston Bay exhibits the largest fluctuations in prediction accuracy. From hour 20 to 28, the lower quartile \(R^2\) in Galveston drops below 0.90, while in Corpus Christi and South Padre Island, the Q1 \(R^2\) generally stays above 0.95 even during peak surge moments. This relative instability in Galveston is likely due to its highly intricate coastal geometry, including numerous narrow estuaries, inland bayous, and artificial channel systems, which pose challenges for convolutional models to resolve. Additionally, Galveston Bay often experiences sharp water elevation changes within small spatial footprints, further challenging pixel-wise regression.

Nevertheless, the overall forecast skill remains strong across all metrics. The worst-case regional RMSE is observed in Galveston at forecast hour 28, reaching 0.035 in normalized RGB space. This corresponds to an average error of 3.5\%, or roughly 9 intensity levels on a standard 8-bit RGB scale. The worst MAE occurs at hour 20 and 24 in Galveston, with a value of 0.019, or approximately 1.9\% pixel error, indicating the model maintains impressive accuracy even during high-surge frames. Despite temporal variability, the median performance in all three regions remains consistently high, confirming the robustness and reliability of the regional models in both average and worst-case scenarios.

\subsubsection{Impact of Physics-Informed Forcing}

Before presenting the combined-region model, we briefly assess the contribution of physics-informed inputs by conducting an ablation study. Specifically, we trained a baseline model that received only the RGB-encoded water elevation inputs, omitting the wind fields and static bathymetry. This control model used the exact same architecture, hidden dimensions, and training hyperparameters as the regional models discussed previously. The only difference was the input dimensionality, which was reduced from six channels to three channels.

\begin{figure}[htbp]
    \centering
    {\small\text{Galveston Bay Example – With Physics-Informed Forcing}}\\
    \includegraphics[width=0.7\textwidth]{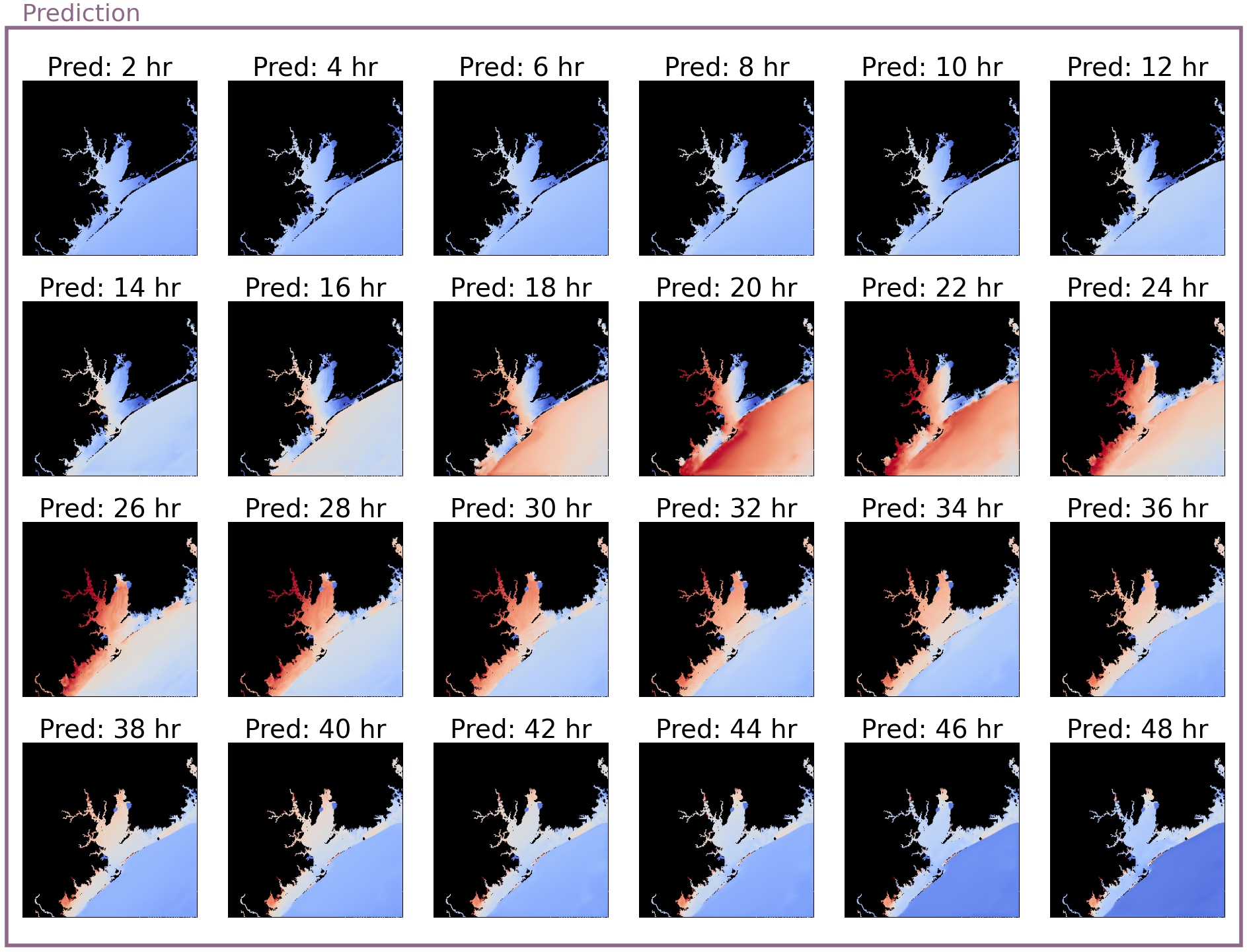}\\[1ex]
    
    {\small\text{Galveston Bay Example – Without Physics-Informed Forcing}}\\
    \includegraphics[width=0.7\textwidth]{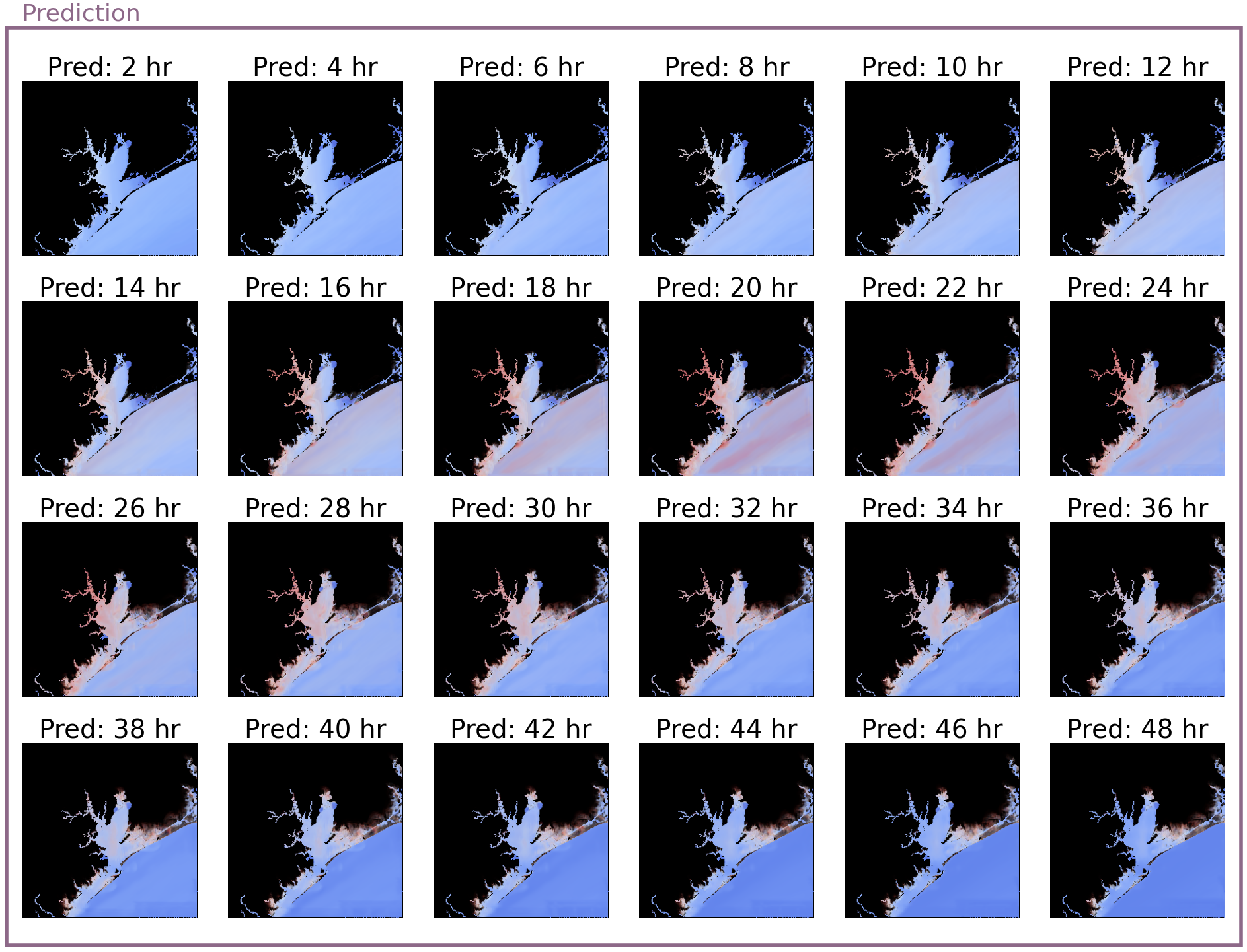}
    
    \caption{Predicted water elevation over time for the same Galveston Bay example shown in Figure~\ref{fig:galveston_example}, using models trained with and without physics-informed forcing. Only predictions are shown here, as the input context and ground truth were provided earlier. Incorporating wind and bathymetry forcings lead to significantly improved spatial and temporal accuracy.}
    
    \label{fig:physics}
\end{figure}

Figure~\ref{fig:physics} compares the predictions from the full physics-informed model and the ablation model for the same Galveston Bay test example shown earlier in Figure~\ref{fig:galveston_example}. Without wind and bathymetric guidance, the model still exhibits some learned intuition of water movement, developed purely from the visual patterns present in the data across thousands of training clips. In particular, it relies on the rate and direction of color gradient changes in the contextual frames to locally extrapolate surge intensity. Its exposure to a large corpus of storm clips allows it to internalize broad temporal patterns, such as how water tends to fade after peak inundation, even in the absence of explicit physical forcing. However, it fails to accurately estimate the magnitude and spatial distribution of inundation, frequently underpredicting surge intensity and misplacing critical flood boundaries.

\subsubsection{Evaluation of Combined-Region Model}

\begin{figure}[htbp]
    \centering
    {\small\text{Cross-Region Prediction Using Regional Models}}\\
    \includegraphics[width=0.7\textwidth]{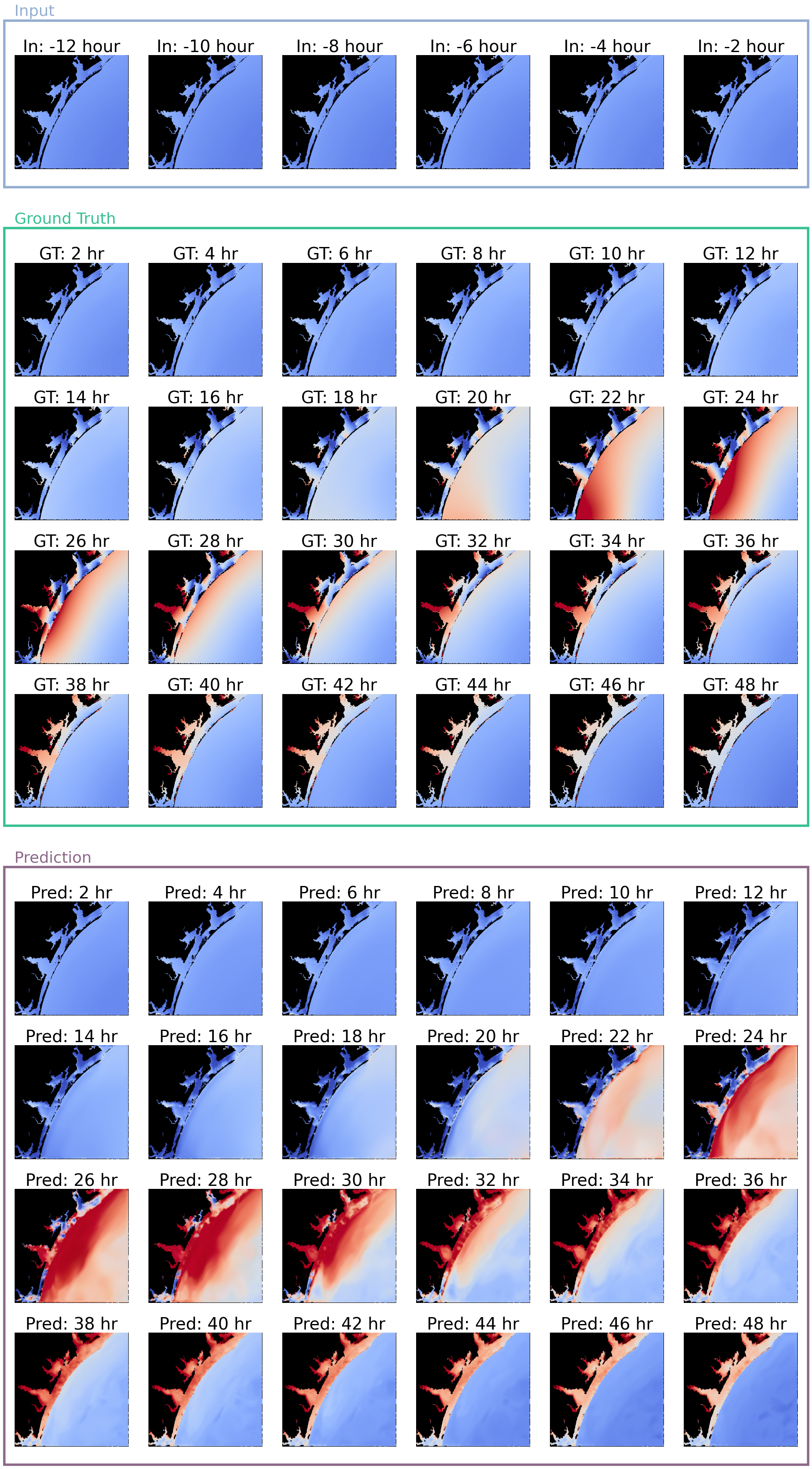}\\[1ex]
    
    \caption{An example of cross-region prediction using the regional models, where a model trained on Galveston Bay is applied to a test clip from the Corpus Christi region. The first box (Input) contains 6 context frames of past water elevation. The second box (Ground Truth) shows the ground truth future evolution, and the third box (Prediction) displays the model's prediction.}
    
    \label{fig:cross_region}
\end{figure}

Regional models are well-suited for evaluating storms that impact specific geographic areas, as they effectively learn the localized interplay between bathymetry, wind forcing, and surge dynamics. However, these models are not ideal for predicting surge in arbitrary or unseen locations. For instance, a model trained solely on Galveston Bay, characterized by its shallow waters and unique coastal geometry, may struggle when applied to regions with markedly different physical characteristics. As illustrated in Figure~\ref{fig:cross_region}, applying the Galveston-trained model to a Corpus Christi test clip results in poor performance: the timing, intensity, and spatial distribution of the surge are all misaligned. Additionally, the model fails to preserve finer coastline details as the forecast progresses, missing complex shoreline dynamics unique to Corpus Christi’s geomorphology. These discrepancies arise because Corpus Christi features deeper bathymetry and different coastline orientation, which alters the frictional interaction between the water column and seabed. Consequently, wind speeds that generate substantial surge in Galveston Bay may be insufficient to produce similar effects in Corpus Christi. These limitations motivate the development of a more generalizable model—one capable of accurately predicting surge onset, intensity, and impact location for previously unseen regions. To achieve this, we combined the training datasets from all three study regions, enabling the model to learn a diverse set of coastal profiles, bathymetric conditions, and wind-surge interactions. The selected regions intentionally vary in coastline shape, orientation, and depth, providing a rich and heterogeneous training that supports broad generalization.

\begin{figure}[htbp]
    \centering
    {\small\text{Combined-Region Model – Full Test Set Performance}}\\
    \includegraphics[width=\textwidth]{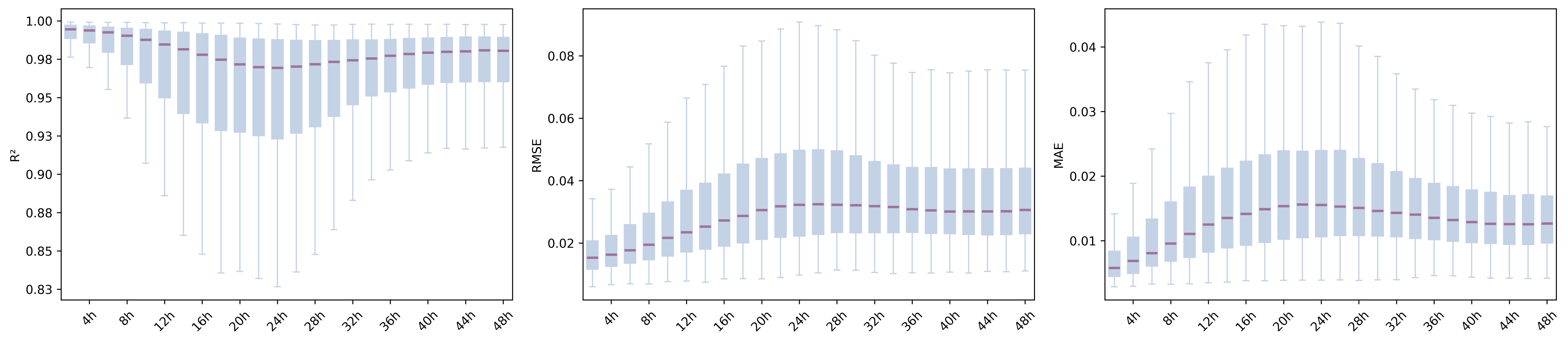}
    
    \caption{Prediction accuracy of the combined-region model evaluated on the full test set, which includes all three regions of interest: Galveston Bay, Corpus Christi, and South Padre Island. Each box shows the distribution of per-frame $R^2$, RMSE, and MAE scores across all test clips at each forecast hour. This visualization reflects the model's overall robustness and generalization across heterogeneous coastal domains.}
    
    \label{fig:whole_general}
\end{figure}

Figure~\ref{fig:whole_general} and Tables~\ref{tab:median_r2}, \ref{tab:median_rmse}, and~\ref{tab:median_mae} present the performance of the combined-region model. The box plots summarize the distribution of per-frame $R^2$, RMSE, and MAE across all test clips from all regions, while the tables provide the corresponding numerical medians. As expected, the overall performance of the combined model is slightly blended between the individual regional models, reflecting its exposure to a broader variety of coastal conditions. The worst-case $R^2$ occurs at forecast hour 24 with a value of 0.969, while the highest RMSE, observed between hours 24 and 32, reaches 0.032 in normalized RGB space. This corresponds to an absolute deviation of approximately 8.2 intensity levels on an 8-bit color scale, a small and acceptable error for pixel-level prediction.

\begin{figure}[htbp]
    \centering
    {\small\text{Regional vs. Combined-Region Model Performance}}\\
    \includegraphics[width=\textwidth]{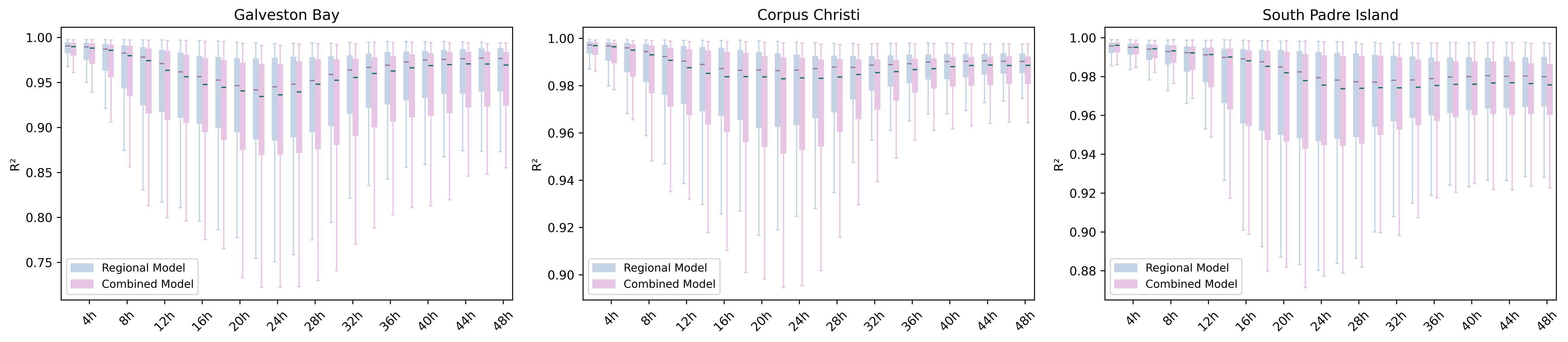}
    \caption{Comparison of $R^2$ performance between region-specific models and the combined-region model, evaluated separately on test clips from Galveston Bay, Corpus Christi, and South Padre Island. Each subplot shows per-frame $R^2$ distributions over 48 hours. The light blue boxes represent performance from regional models, while the pink boxes represent the same region’s results from the combined-region model.}
    \label{fig:regional_vs_whole}
\end{figure}

To assess the trade-off between specialization and generalization, we compare the combined model's performance on each region’s test set against its corresponding regional model, as shown in Figure~\ref{fig:regional_vs_whole}. Across all three regions, the combined model shows a minor, consistent degradation in forecast accuracy at nearly every hour, with slightly lower $R^2$ and marginally higher RMSE and MAE. However, this small performance trade-off is offset by a major benefit: as demonstrated in the next section, the combined model significantly outperforms regional models in extrapolating to out-of-distribution regions, delivering robust predictions for previously unseen coastal geometries.

\begin{table}[htbp]
\centering
\small
\caption{Median $R^2$ scores at selected forecast hours.}
\label{tab:median_r2}
\begin{tabular}{lcccccccccccc}
\toprule
Region & 4h & 8h & 12h & 16h & 20h & 24h & 28h & 32h & 36h & 40h & 44h & 48h \\
\midrule
Galveston       & 0.989 & 0.983 & 0.971 & 0.956 & 0.946 & 0.945 & 0.952 & 0.964 & 0.969 & 0.975 & 0.976 & 0.977 \\
Corpus Christi  & 0.997 & 0.994 & 0.990 & 0.987 & 0.987 & 0.987 & 0.988 & 0.989 & 0.989 & 0.990 & 0.990 & 0.990 \\
South Padre     & 0.995 & 0.993 & 0.991 & 0.989 & 0.985 & 0.979 & 0.977 & 0.978 & 0.979 & 0.980 & 0.980 & 0.980 \\
Combined Model  & 0.994 & 0.990 & 0.985 & 0.978 & 0.972 & 0.969 & 0.972 & 0.974 & 0.977 & 0.979 & 0.980 & 0.981 \\
\bottomrule
\end{tabular}
\end{table}

\begin{table}[htbp]
\centering
\small
\caption{Median RMSE values at selected forecast hours.}
\label{tab:median_rmse}
\begin{tabular}{lcccccccccccc}
\toprule
Region & 4h & 8h & 12h & 16h & 20h & 24h & 28h & 32h & 36h & 40h & 44h & 48h \\
\midrule
Galveston       & 0.019 & 0.022 & 0.026 & 0.030 & 0.032 & 0.034 & 0.035 & 0.034 & 0.033 & 0.033 & 0.032 & 0.033 \\
Corpus Christi  & 0.012 & 0.016 & 0.020 & 0.023 & 0.024 & 0.024 & 0.024 & 0.023 & 0.022 & 0.022 & 0.022 & 0.022 \\
South Padre     & 0.015 & 0.018 & 0.020 & 0.022 & 0.026 & 0.030 & 0.032 & 0.031 & 0.031 & 0.031 & 0.031 & 0.031 \\
Combined Model  & 0.016 & 0.019 & 0.023 & 0.027 & 0.031 & 0.032 & 0.032 & 0.032 & 0.031 & 0.030 & 0.030 & 0.031 \\
\bottomrule
\end{tabular}
\end{table}

\begin{table}[htbp]
\centering
\small
\caption{Median MAE values at selected forecast hours.}
\label{tab:median_mae}
\begin{tabular}{lcccccccccccc}
\toprule
Region & 4h & 8h & 12h & 16h & 20h & 24h & 28h & 32h & 36h & 40h & 44h & 48h \\
\midrule
Galveston       & 0.008 & 0.010 & 0.014 & 0.016 & 0.019 & 0.019 & 0.018 & 0.016 & 0.015 & 0.014 & 0.014 & 0.014 \\
Corpus Christi  & 0.005 & 0.008 & 0.011 & 0.012 & 0.012 & 0.011 & 0.011 & 0.010 & 0.010 & 0.009 & 0.009 & 0.009 \\
South Padre     & 0.006 & 0.009 & 0.011 & 0.012 & 0.013 & 0.013 & 0.013 & 0.012 & 0.012 & 0.011 & 0.011 & 0.011 \\
Combined Model  & 0.007 & 0.010 & 0.013 & 0.014 & 0.015 & 0.016 & 0.015 & 0.014 & 0.014 & 0.013 & 0.013 & 0.013 \\
\bottomrule
\end{tabular}
\end{table}

\clearpage
\subsection{Out-of-Distribution Generalization}

In this section, we evaluate the generalization capability of the combined-region model by testing its performance on out-of-distribution regions. Specifically, we applied the model to coastal areas that were not included in the training set, assessing its ability to extrapolate surge dynamics in previously unseen geographic contexts. We also evaluated its performance on historical high-impact events, such as Hurricane Ike, to better understand its behavior under real-world storm conditions.

\subsubsection{Testing on Geographically Unseen Regions}

\begin{figure}[h]
    \centering
    \includegraphics[width=\textwidth]{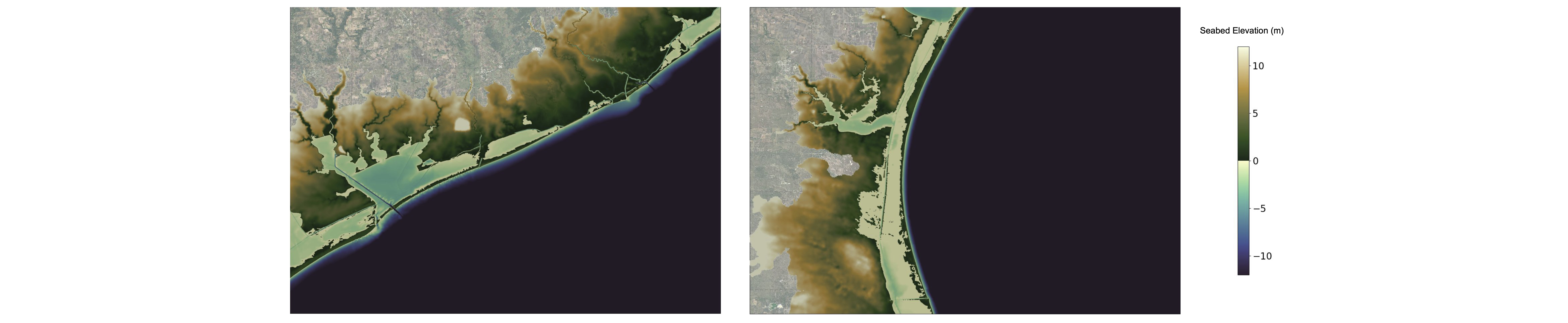}
    \caption{Bathymetric structure of the two out-of-distribution test regions: Matagorda Bay and Baffin Bay.}
    \label{fig:bathy_extra}
\end{figure}

We defined two geographically unseen regions for evaluating out-of-distribution generalization: Matagorda Bay and Baffin Bay. These regions are challenging due to their complex coastal geometries and detailed bathymetric features, which demand strong spatial generalization from the model. Matagorda Bay is situated between Galveston Bay and Corpus Christi, sharing the intricate inner-bay coastline of Galveston and the deeper offshore waters characteristic of Corpus Christi. Baffin Bay lies between Corpus Christi and South Padre Island, with a coastal geometry similar to Corpus Christi and an open-sea orientation resembling that of South Padre. Figure~\ref{fig:bathy_extra} illustrates the bathymetric structures of these two out-of-distribution regions.

\begin{figure}[htbp]
    \centering
    {\small\text{Prediction Visualization for a Single Matagorda Bay Example}}\\
    \includegraphics[width=0.7\textwidth]{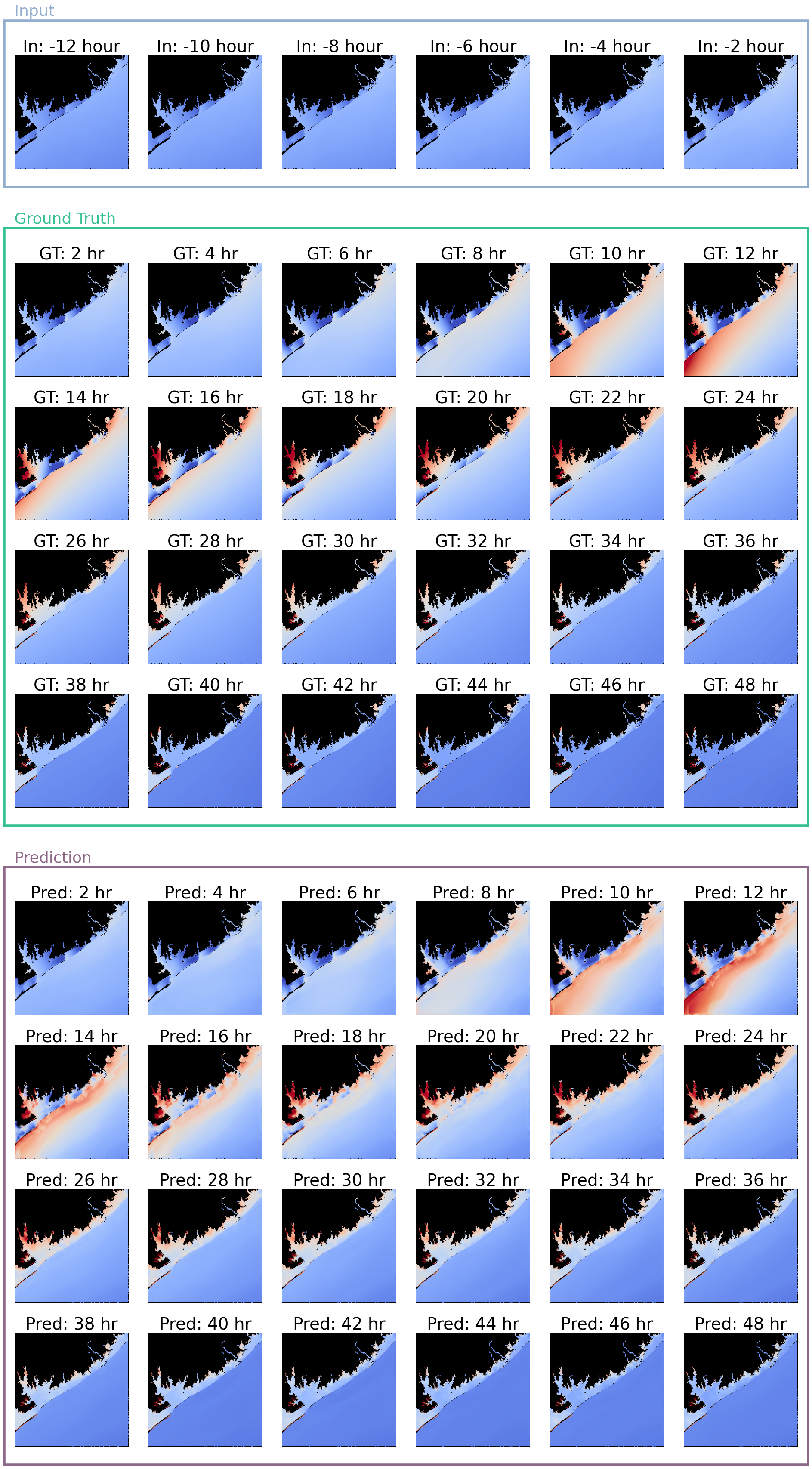}\\
    
    \caption{A single example from Matagorda Bay, a region not included during training. The first box (Input) contains 6 context frames of past water elevation. The second box (Ground Truth) shows the ground truth future evolution, and the third box (Prediction) displays the model's prediction.}
    
    \label{fig:matagorda_example}
\end{figure}

\begin{figure}[htbp]
    \centering
    
    {\small\text{Prediction Accuracy Over Forecast Time}}\\
    \includegraphics[width=\textwidth]{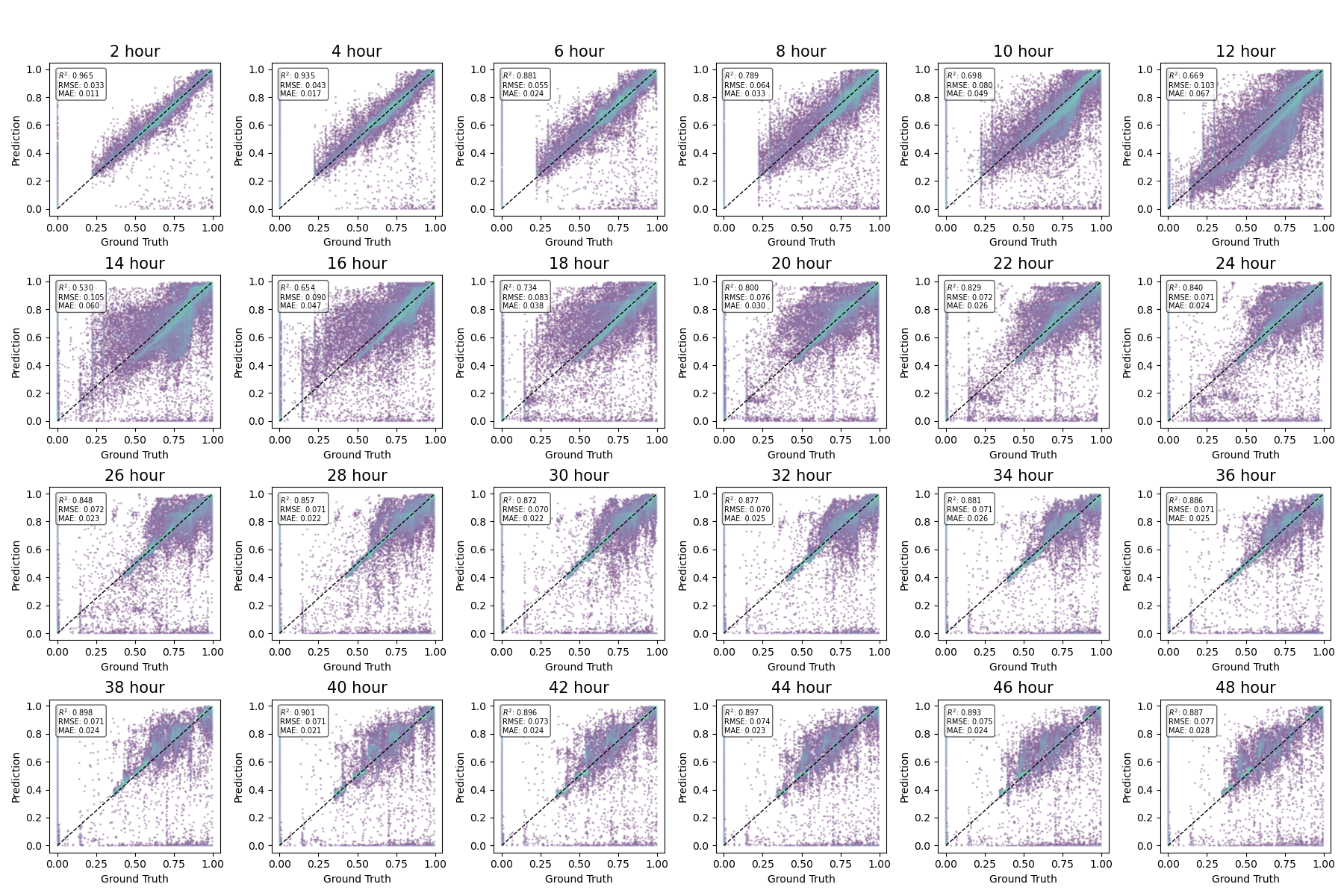}
    
    \caption{Prediction accuracy metrics ($R^2$, RMSE, and MAE) at each forecast hour for the same Matagorda Bay example shown in Figure~\ref{fig:matagorda_example}, reflecting the model’s generalization ability to out-of-distribution coastal geography.}
    
    \label{fig:matagorda_metrics}
\end{figure}

\begin{figure}[htbp]
    \centering
    {\small\text{Prediction Visualization for a Single Baffin Bay Example}}\\
    \includegraphics[width=0.7\textwidth]{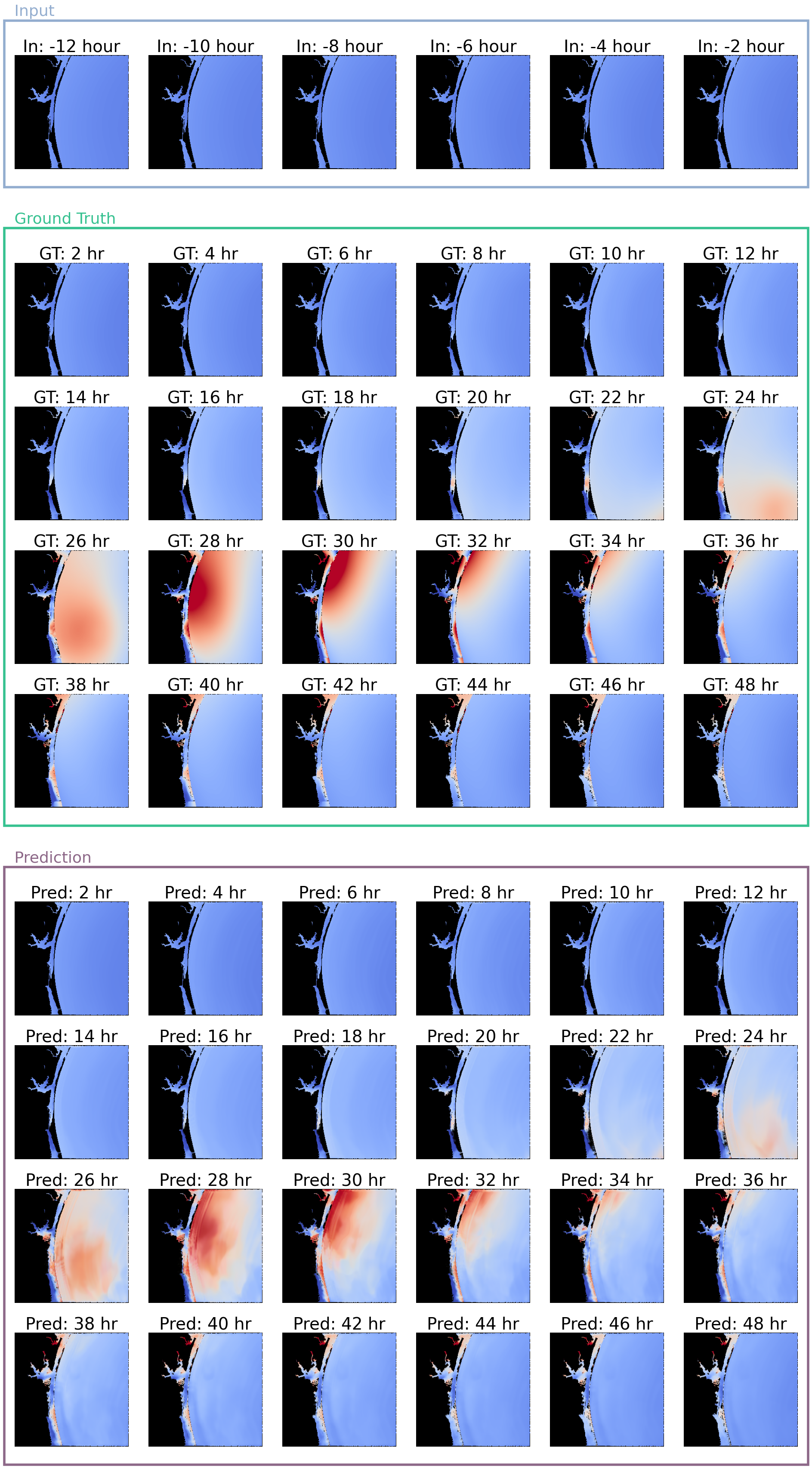}\\
    
    \caption{A single example from Baffin Bay, a region not included during training. The first box (Input) contains 6 context frames of past water elevation. The second box (Ground Truth) shows the ground truth future evolution, and the third box (Prediction) displays the model's prediction.}
    
    \label{fig:baffin_example}
\end{figure}

\begin{figure}[htbp]
    \centering

    {\small\text{Prediction Accuracy Over Forecast Time}}\\
    \includegraphics[width=\textwidth]{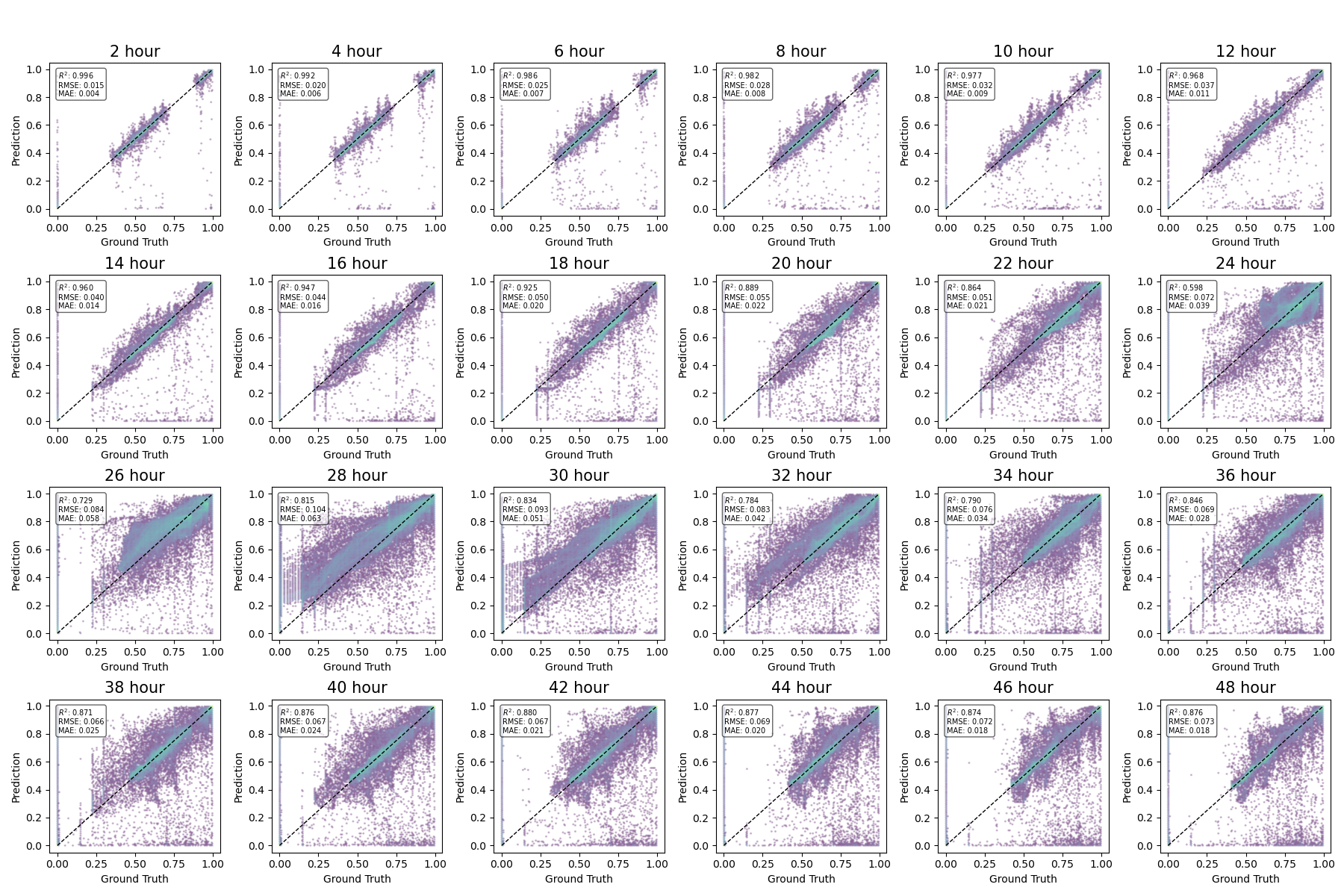}
    
    \caption{Prediction accuracy metrics ($R^2$, RMSE, and MAE) at each forecast hour for the same Baffin Bay example shown in Figure~\ref{fig:baffin_example}, reflecting the model’s generalization ability to out-of-distribution coastal geography.}
    
    \label{fig:baffin_metrics}
\end{figure}

The model demonstrates remarkable generalization to previously unseen coastal regions. Figures~\ref{fig:matagorda_example} and~\ref{fig:baffin_example} show representative test clips from Matagorda Bay and Baffin Bay, respectively, where the visualized prediction over time is presented. In each figure, the first box contains contextual input, the second box shows the ground truth surge progression, and the third box displays the model predictions. Corresponding accuracy metrics for each example are shown in Figures~\ref{fig:matagorda_metrics} and~\ref{fig:baffin_metrics}, which provide per-frame performance metrics ($R^2$, RMSE, and MAE) over the entire prediction sequence.

Visually, the results are striking. For both regions, the model correctly predicts the onset timing of surge, peak inundation levels, spatial distribution of water along the coast, and the gradual retreat of water following the surge. Particularly impressive is the model's ability to resolve detailed inundation patterns in small rivers, estuaries, and partially submerged landforms. The predicted geometry of temporary flooding along narrow floodplains and coastal islands matches ground truth closely, even though these regions were not part of the training data. This suggests that the model is not merely memorizing regional coastline shapes, but has successfully learned transferable surge dynamics grounded in the physical interplay between wind forcing, bathymetry, and coastal morphology.

The quantitative plots in Figures~\ref{fig:matagorda_metrics} and~\ref{fig:baffin_metrics} support these visual observations. For the Matagorda Bay example, the worst frame-level performance occurs at forecast hour 14 with $R^2 = 0.530$, while Baffin Bay's lowest frame-level score is $R^2 = 0.598$ at hour 24. Despite these isolated dips, the majority of frames in both examples achieve $R^2 > 0.8$, indicating strong overall temporal stability and predictive power.

\begin{figure}[htbp]
    \centering
    {\small\text{Matagorda Bay - Regional Performance Overview}}\\
    \includegraphics[width=\textwidth]{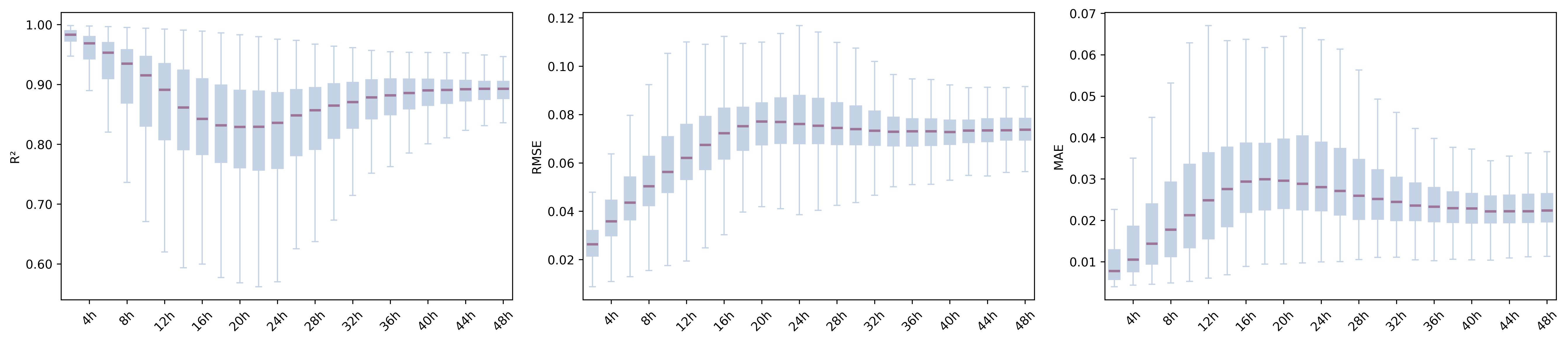}
    
    \caption{Prediction accuracy across all test clips in the Matagorda Bay region. Each box represents the distribution of per-frame $R^2$, RMSE, and MAE scores computed across all clips at each forecast hour. Compared to the single-clip result, this visualization summarizes the model's generalization performance across the entire region.}
    
    \label{fig:matagorda_general}
\end{figure}

\begin{figure}[htbp]
    \centering
    {\small\text{Baffin Bay - Regional Performance Overview}}\\
    \includegraphics[width=\textwidth]{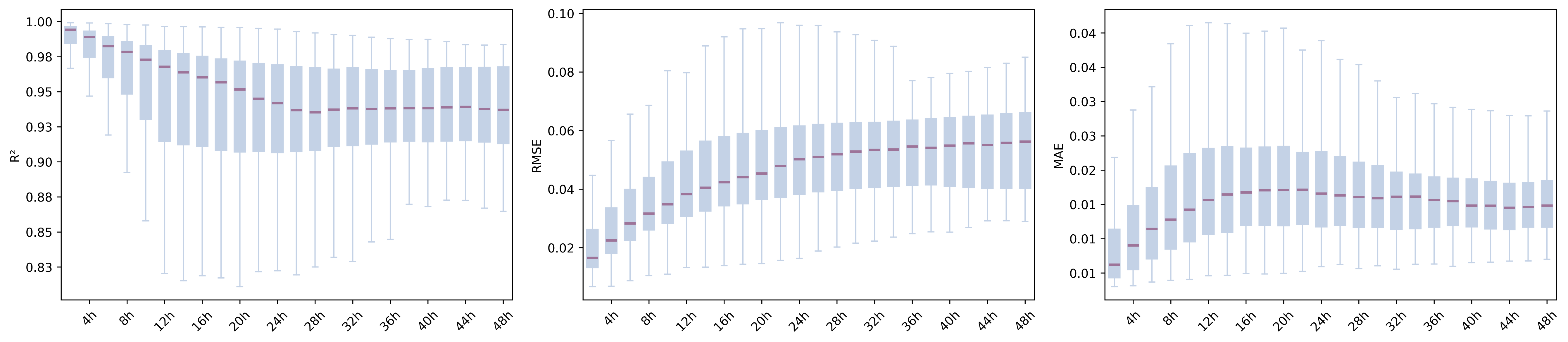}
    
    \caption{Prediction accuracy across all test clips in the Baffin Bay region. Each box represents the distribution of per-frame $R^2$, RMSE, and MAE scores computed across all clips at each forecast hour. Compared to the single-clip result, this visualization summarizes the model's generalization performance across the entire region.}
    
    \label{fig:baffin_general}
\end{figure}

Figures~\ref{fig:matagorda_general} and~\ref{fig:baffin_general}, along with Tables~\ref{tab:median_r2_extra},~\ref{tab:median_rmse_extra}, and~\ref{tab:median_mae_extra}, present the aggregate accuracy metrics over all test clips in the two regions. In Matagorda Bay, the lowest median $R^2$ occurs at forecast hour 20 with a value of 0.829, and the worst RMSE also at hour 20 is 0.077, corresponding to an average absolute pixel deviation of approximately 20 units in 8-bit RGB space. For Baffin Bay, the lowest $R^2$ is 0.935 at hour 28, and the highest RMSE is 0.056 at hour 48, about 14 RGB intensity levels. Importantly, median $R^2$ values remain above 0.8 for Matagorda and above 0.9 for Baffin Bay across all forecast frames, demonstrating robust generalization with minimal degradation in performance. These results underscore the model's ability to extrapolate reliably beyond the training regions while preserving spatial and temporal coherence.

\begin{table}[htbp]
\centering
\small
\caption{Median $R^2$ scores at selected forecast hours in out-of-distribution regions.}
\label{tab:median_r2_extra}
\begin{tabular}{lcccccccccccc}
\toprule
Region & 4h & 8h & 12h & 16h & 20h & 24h & 28h & 32h & 36h & 40h & 44h & 48h \\
\midrule
Matagorda Bay & 0.969 & 0.935 & 0.891 & 0.843 & 0.829 & 0.836 & 0.857 & 0.871 & 0.882 & 0.890 & 0.892 & 0.893 \\
Baffin Bay    & 0.989 & 0.978 & 0.968 & 0.960 & 0.952 & 0.942 & 0.935 & 0.938 & 0.938 & 0.938 & 0.939 & 0.937 \\
\bottomrule
\end{tabular}
\end{table}

\begin{table}[htbp]
\centering
\small
\caption{Median RMSE values at selected forecast hours in out-of-distribution regions.}
\label{tab:median_rmse_extra}
\begin{tabular}{lcccccccccccc}
\toprule
Region & 4h & 8h & 12h & 16h & 20h & 24h & 28h & 32h & 36h & 40h & 44h & 48h \\
\midrule
Matagorda Bay & 0.036 & 0.050 & 0.062 & 0.072 & 0.077 & 0.076 & 0.075 & 0.073 & 0.073 & 0.073 & 0.073 & 0.074 \\
Baffin Bay    & 0.023 & 0.032 & 0.038 & 0.042 & 0.045 & 0.050 & 0.052 & 0.053 & 0.055 & 0.055 & 0.055 & 0.056 \\
\bottomrule
\end{tabular}
\end{table}

\begin{table}[htbp]
\centering
\small
\caption{Median MAE values at selected forecast hours in out-of-distribution regions.}
\label{tab:median_mae_extra}
\begin{tabular}{lcccccccccccc}
\toprule
Region & 4h & 8h & 12h & 16h & 20h & 24h & 28h & 32h & 36h & 40h & 44h & 48h \\
\midrule
Matagorda Bay & 0.011 & 0.018 & 0.025 & 0.029 & 0.030 & 0.028 & 0.026 & 0.024 & 0.023 & 0.023 & 0.022 & 0.022 \\
Baffin Bay    & 0.009 & 0.013 & 0.016 & 0.017 & 0.017 & 0.017 & 0.016 & 0.016 & 0.016 & 0.015 & 0.015 & 0.015 \\
\bottomrule
\end{tabular}
\end{table}

\subsubsection{Validation on Historical Storm Events}

To further evaluate the model's practical utility, we tested its performance on a real-world extreme event, Hurricane Ike, a Category 4 major hurricane that formed in early September 2008 and caused catastrophic damage along the Texas Gulf Coast. With sustained winds reaching 145 mph and over \$38 billion in damages, Hurricane Ike stands as one of the most destructive hurricanes in U.S. history. Importantly, it provides a unique opportunity to assess the model’s robustness under conditions that are well beyond the scope of the training surge distribution.

\begin{figure}[htbp!]
    \centering
    {\small\text{Prediction Visualization for Hurricane Ike in Galveston Bay}}\\
    \includegraphics[width=0.7\textwidth]{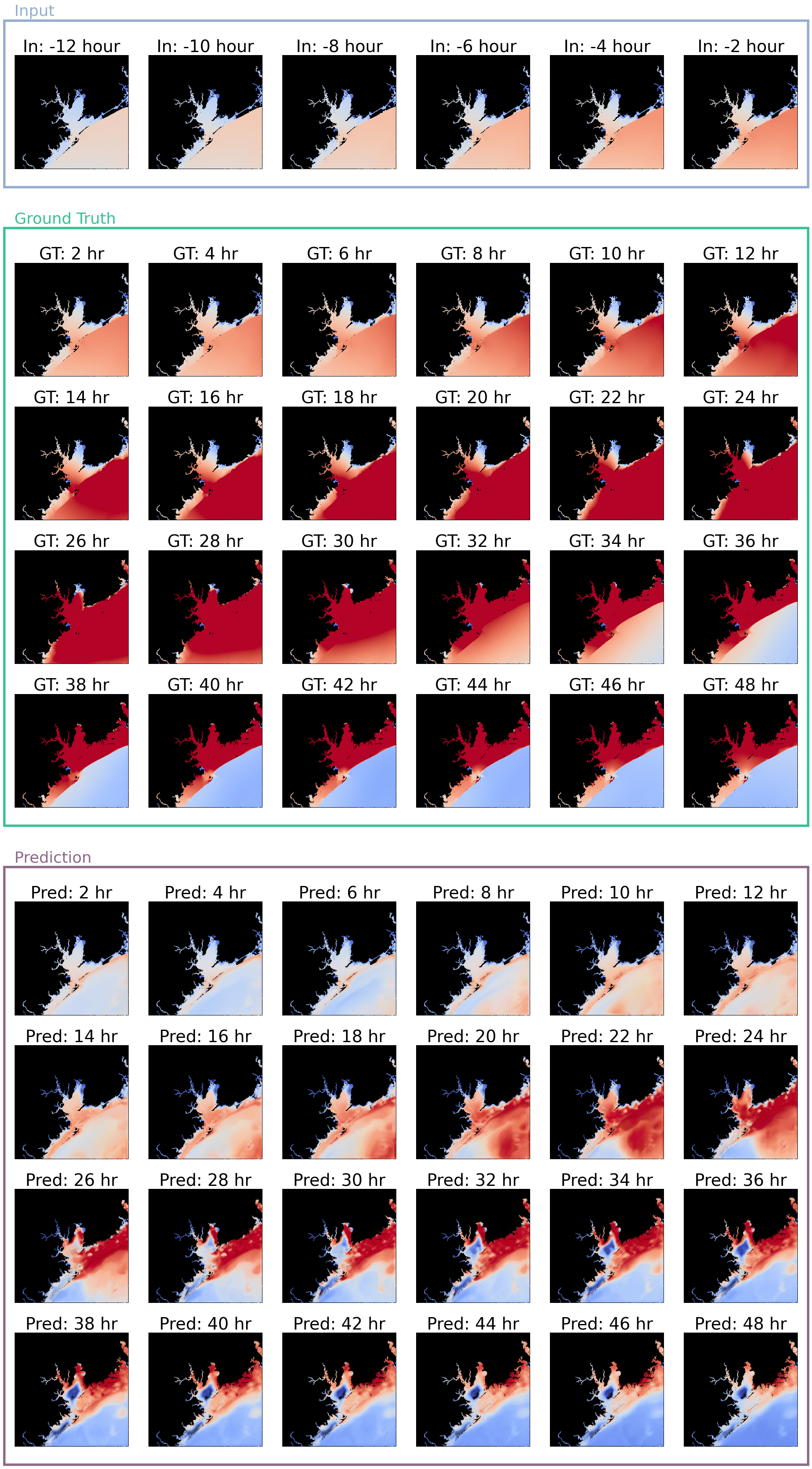}
    \caption{Prediction visualization for Hurricane Ike (2008) in Galveston Bay using the combined-region model. Due to extreme surge levels well beyond the model’s training distribution, the prediction significantly underestimates both the intensity and spatial extent of inundation.}
    \label{fig:galveston_ike}
\end{figure}

We evaluated the combined-region model on two coastal locations affected by Hurricane Ike: Galveston Bay and Corpus Christi. As shown in Figure~\ref{fig:galveston_ike}, the model is able to capture the general spatial pattern and progression of the surge in Galveston Bay, but significantly underpredicts the magnitude and extent of inundation. This degradation is expected since surge levels during Hurricane Ike in Galveston Bay exceeded 5 meters in many areas, whereas the training data were normalized with a maximum range of 2.5 meters. The model simply has not seen dynamics of this scale, making it ill-equipped to predict such extreme outcomes.

\begin{figure}[htbp!]
    \centering
    {\small\text{Prediction Visualization for Hurricane Ike in Corpus Christi}}\\
    \includegraphics[width=0.7\textwidth]{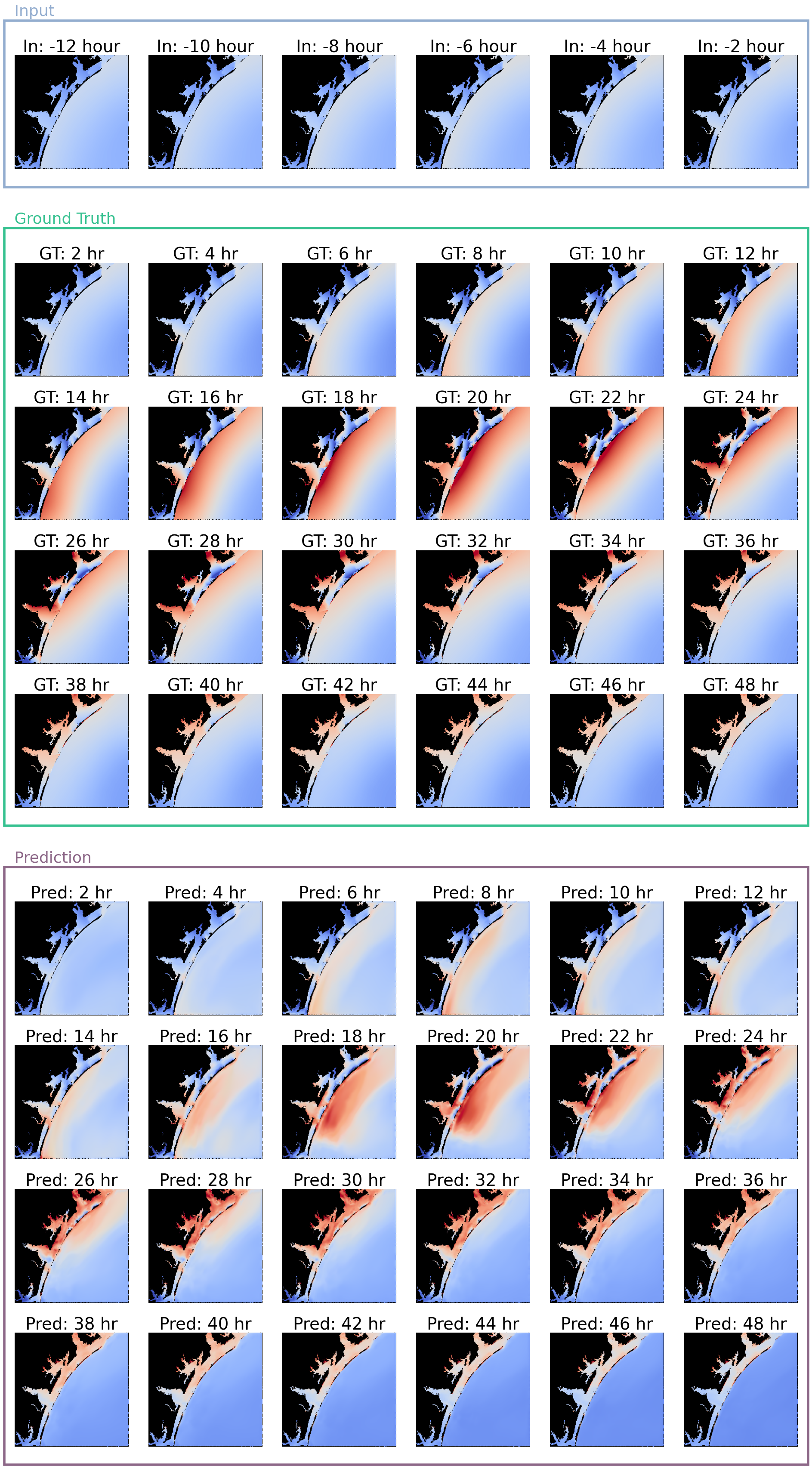}
    \caption{Prediction visualization for Hurricane Ike (2008) in Corpus Christi using the combined-region model. Here, surge magnitudes mostly remain within the model’s training range, resulting in accurate spatial and temporal predictions.}
    \label{fig:corpus_ike}
\end{figure}

\begin{figure}[t!]
    \centering
    \includegraphics[width=0.5\textwidth]{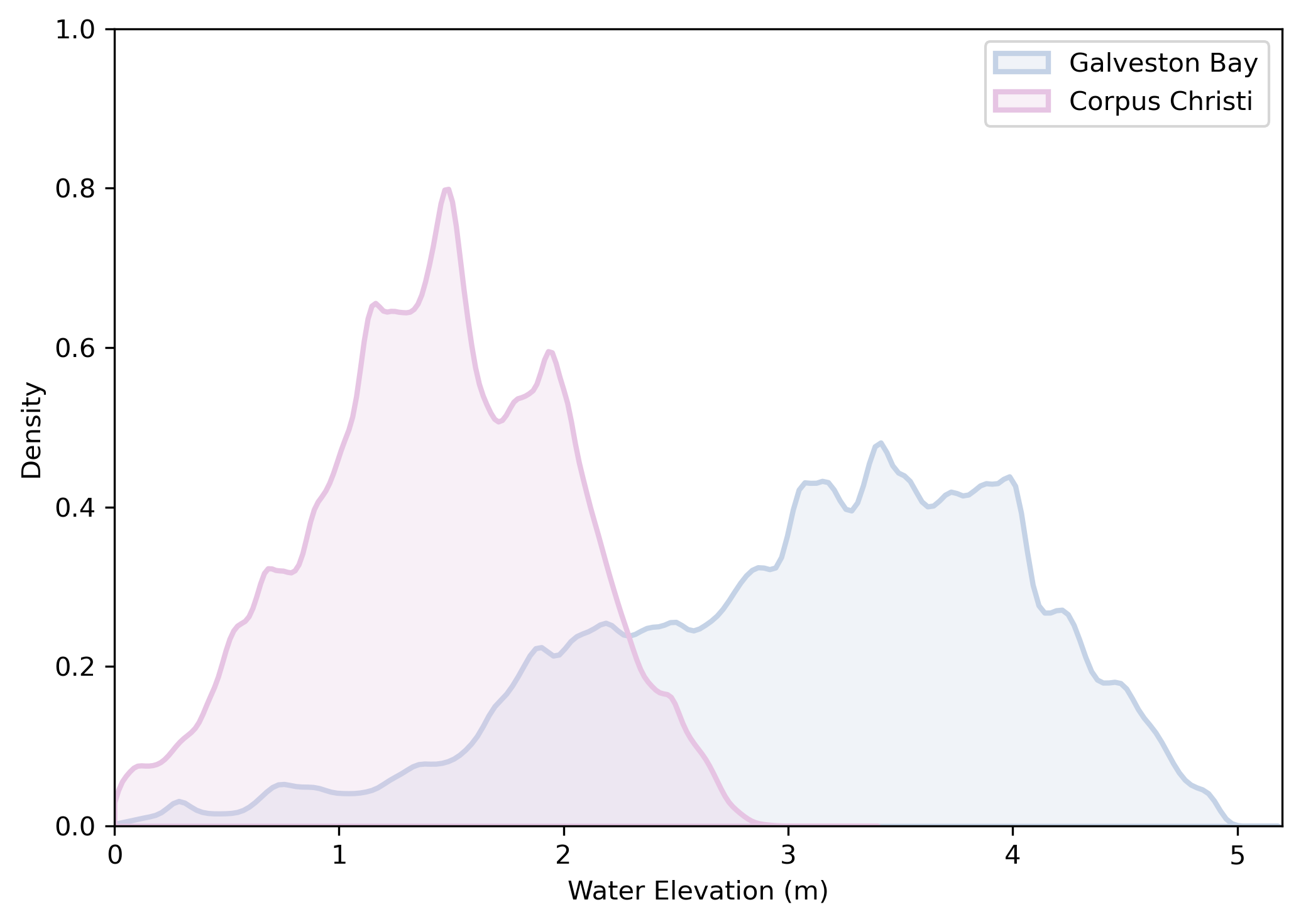}
    \caption{Water elevation distributions during the peak 12-hour window for Galveston Bay and Corpus Christi during Hurricane Ike (2008). Galveston exhibits significantly higher surge levels, exceeding the normalized range of the training distribution.}
    \label{fig:ike_distribution}
\end{figure}

In contrast, Figure~\ref{fig:corpus_ike} shows the model’s prediction in Corpus Christi during the same event. Here, surge magnitudes mostly remain within the training surge distribution’s range, and the model delivers accurate predictions in terms of timing, intensity, and spatial location of flooding. 
This contrast highlights a key limitation of the current approach: while the model generalizes well across different spatial regions and temporal windows, its numerical extrapolation capability—especially for rare, high-magnitude events—is more constrained.
To better illustrate this, Figure~\ref{fig:ike_distribution} shows the water elevation distributions during the peak 12-hour window in both regions. In Corpus Christi, the surge values are well contained within the training normalization range, mostly below 2.5 meters. In Galveston Bay, however, the majority of values exceed 3 meters, with many reaching as high as 5 meters. 
This stark difference underscores the challenge of forecasting extreme events when the target values fall well outside the training distribution. We discuss potential directions to overcome this limitation in the concluding section.

\section{Conclusions and Future Work}
\label{sec:conclusion}

This work presents a novel and practical approach to storm surge forecasting by leveraging RGB-encoded water elevation maps and physics-aware deep learning. By encoding spatial temporal surge evolution into RGB image sequences, we enable the use of ConvLSTM models for autoregressive forecasting, treating storm surge prediction as a video prediction task. Our method is trained on a large synthetic dataset generated by ADCIRC simulation outputs, capturing thousands of storms clips along the Texas coast. The results, as discussed in Section~\ref{sec:result}, demonstrate highly robust performance across multiple regions, with median $R^2$ values consistently exceeding 0.9 for most forecast hours. More remarkably, the model exhibits strong generalization to geographically unseen regions, capturing complex inundation patterns and surge propagation in areas never seen during training, without relying on explicit physical equations or simulation feedback.

However, the current model has limitations. Its predictive capability is bounded by the surge distribution observed in the training data, which limits its robustness when encountering extreme events. For example, as demonstrated in Section~\ref{sec:result}, Hurricane Ike (2008) produced peak water elevations exceeding 5 meters in Galveston Bay, far beyond the range observed during training, leading to significant degradation in predictive accuracy. While the model demonstrates strong generalization across space and time by successfully predicting surge in unseen geographic regions and future time steps, it remains limited in its ability to extrapolate to numerically extreme events.

With the current work serving as a benchmark, future efforts will focus on scaling both the dataset and model architecture to improve generalization and robustness.
This includes expanding spatially, temporally, and numerically by incorporating more coastal regions with diverse conditions and a broader range of storm intensities and surge magnitudes. Architecturally, while ConvLSTM offers a strong baseline, it is not well suited for extremely large datasets due to its limited memory capacity and inherently sequential architecture, which restricts parallelization and leads to prohibitively long training times. To address this, we plan to explore advanced models such as attention-based transformers, diffusion models, and memory-augmented recurrent architectures that can better capture long term dependencies and high dimensional dynamics. Another avenue for future research is probabilistic forecasting. The current ConvLSTM is deterministic and autoregressive, generating one frame at a time based solely on the immediate past. Future models should instead estimate distributions over possible outcomes using generative frameworks such as diffusion-based video prediction and probabilistic sequence modeling. These improvements will enable more informative, uncertainty aware forecasts that reflect a range of possible outcomes and support effective storm surge response planning.

In conclusion, this work demonstrates a successful application of RGB-encoded storm surge forecasting, integrating coastal geometry and physical forcing in a compact, learnable format. Our ConvLSTM-based framework achieves strong performance across seen regions and generalizes well to unseen coastlines, highlighting the promise of this approach. Looking ahead, future work will explore larger and more diverse datasets, more scalable architectures, and probabilistic forecasting to further improve robustness and practical applicability.

\section*{Code Availability}
All code and scripts used to train and evaluate the model are publicly available at \url{https://github.com/MaxPaiPai/Storm_Surge_in_Color}.

\section*{Acknowledgments}
This work was supported by a subcontract from  Los Alamos National Laboratory under Subcontract C5223 and Master Agreement BA\# 588340.

The authors also would like to gratefully acknowledge the use of the "ADCIRC" allocation on the Vista supercomputer, and ``ADCIRC", ``DMS23001", and ``DMS21031" allocations on the Frontera supercomputer at The Texas Advanced Computing Center at The University of Texas at Austin.

\clearpage

\bibliography{references}

\end{document}